\documentclass[twocolumn,         
               showpacs,            
               preprintnumbers,     
               aps,                 
               prd,          	    
               letterpaper,             
               superscriptaddress,      
               nofootinbib,         
               tightenlines,        
               floats,floatfix      
               ]{revtex4-1}
\usepackage{dcolumn}   
\usepackage{bm}        
\usepackage{float}

\usepackage{graphicx,color}
\usepackage{latexsym}
\usepackage{amsmath,amssymb}        
\usepackage[colorlinks=true,linkcolor=blue,citecolor=blue]{hyperref}
\usepackage{mathrsfs}
\usepackage{comment}
\usepackage{soul}
\definecolor{purple}{rgb}{0.58,0.0,0.83}
\definecolor{orange}{rgb}{1,0.5,0}
\DeclareSymbolFontAlphabet{\mathrsfs}{rsfs}
\DeclareMathAlphabet{\mathcal}{OMS}{cmsy}{m}{n}

\begin{document}


\title{Gravitational atoms: general framework for the construction of multistate axially symmetric solutions of the Schr\"odinger-Poisson system}


\author{F. S. Guzm\'an}
\email{francisco.s.guzman@umich.mx}
\affiliation{Laboratorio de Inteligencia Artificial y Superc\'omputo,
	      Instituto de F\'{\i}sica y Matem\'{a}ticas, Universidad
              Michoacana de San Nicol\'as de Hidalgo. Edificio C-3, Cd.
              Universitaria, 58040 Morelia, Michoac\'{a}n,
              M\'{e}xico.}
\author{L. Arturo Ure\~{n}a-L\'{o}pez} 
 \email{lurena@ugto.mx}
\affiliation{%
Departamento de F\'isica, DCI, Campus Le\'on, Universidad de
Guanajuato, 37150, Le\'on, Guanajuato, M\'exico.}


\date{\today}


\begin{abstract}
We present a general strategy to solve the stationary Schr\"odinger-Poisson (SP) system of equations for multistates with axial symmetry. The approach allows us to obtain the well known single and multistate solutions with spherical symmetry, Newtonian multistate $\ell-$boson stars and axially symmetric multistate configurations. For each case we construct particular examples that illustrate the method, whose stability properties are studied by numerically solving the time-dependent SP system. Among the stable configurations there are the mixed-two-state configurations including spherical and dipolar components, which might have an important value as potential anisotropic dark matter halos in the context of ultralight bosonic dark matter scenarios. This is the reason why we also present a possible process of formation of these mixed-two-state configurations that could open the door to the exploration of more general multistate structure formation scenarios.
\end{abstract}


\pacs{keywords: self-gravitating systems -- dark matter -- Bose condensates}


\maketitle

Systems of self-gravitating scalar bosons have been widely studied and discussed ever since the appearance of the seminal work in Ref.~\cite{Ruffini:1969qy}. The main feature being the existence of stable equilibrium configurations, which is the result of a well-posed eigenvalue problem of the Einstein-Klein-Gordon system of equations. The stability of such systems has been studied by both, semi-analytical and numerical means, see for instance the comprehensive reviews~\cite{Schunck:2003kk,Liebling:2012fv} and references therein. Although most of the studies have focused on spherically symmetric configurations, there are already some of them that have tested the properties of the scalar field systems under general circumstances, e. g. axially-symmetric or rotating solutions, see~\cite{Liebling:2012fv,Mielke:2016war}, including cases in the Newtonian regime \cite{1996PhLB..366...85S}.

Apart from  applications in astrophysical situations involving compact objects, there was a renewed interest in self-gravitating bosons because of their possible role as dark matter candidates in the galactic and cosmological contexts~\cite{Matos:2008ag,Suarez:2013iw,Marsh:2015xka,Hui:2016ltb,Urena-Lopez:2019kud}. In particular, the appropriate setup for the formation of galaxies is the non-relativistic, Newtonian regime, conditions under which the Einstein-Klein-Gordon equations become the so-called Schr\"odinger-Poisson (SP) system \cite{Seidel:1990jh,Guzman:2004wj}. The SP system rules the dynamics of ultralight bosonic dark matter, and under the assumption that bosons are ultralight with masses of order $m_a \simeq 10^{-22} \, \mathrm{eV}/c^2$ and occupy  one state, the model shows spectacular advances, including  structure formation simulations indicating attractor density profiles of structures associated to this model \cite{Hu:2000ke,Sahni:1999qe,Matos:2000ng,Hu:2000ke,Schive:2014hza,Mocz:2017wlg,Mocz:2019uyd}.  In fact, state of the art structure formation simulations with this dark matter now can track the formation of axial structures, their further dynamics and its interaction with baryonic dark matter \cite{2019PhRvL.123n1301M}.

As suggested already in Ref.~\cite{Ruffini:1969qy}, there is the possibility to consider the population of different eigenstates in a system of self-gravitating bosons, an idea that was in turn taken for the construction of more general configurations to model galaxy dark matter halos in~\cite{Matos:2007zza}. This results in the so-called multistate system that was studied rigorously in~\cite{Bernal:2009zy,UrenaLopez:2010ur}, where it was found that the system was gravitationally stable as long as the ground state is the most populated one. However, the studies have focused on spherical multistate systems, and are evolving toward more general scenarios \cite{Li:2019mlk}), which is a door open that can expand the scientific potential of the boson dark matter model.

In this manuscript, we present a general approach for the construction of equilibrium configurations with mixed states of the SP system of equations with axial symmetry. For that we follow guidance from previous works, specially for the chosen ansatzs of the scalar wavefunction and the gravitational potential~\cite{Silveira:1995dh,Hertzberg:2018lmt,Davidson:2016uok,Li:2019mlk}. The resultant combination of various states resembles the structure of the electronic cloud in atoms. 

\textit{General framework.} The SP system of equations, in variables absorbing the constants $\hbar$, $G$, the boson mass $m_a$ and without self-interaction, for a combination of ortogonal states $n\ell m$ is

\begin{eqnarray}
i\dot{\Psi}_{n \ell m} &=& -\frac{1}{2}\nabla^2 \Psi_{n \ell m} + V\Psi_{n\ell m} \, , \; \nabla^2 V = \sum_{n \ell m}|\Psi_{n \ell m} |^2 \, . \label{eq:gpp}
\end{eqnarray}

\noindent Here, $\Psi_{n \ell m}=\Psi_{n \ell m}(t,{\bf x})$ is an order parameter describing the macroscopic behavior of the boson gas, so that $|\Psi_{n \ell m}|^2$ is the mass density of the given state, and $V=V(t,{\bf x})$ is the gravitational potential sourced by the bosonic clouds in different states. 

We assume the following ansatz for the wave function

\begin{equation}
    \Psi_{n\ell m}(t,{\bf x}) = \sqrt{4\pi} \, e^{-i\gamma_{n\ell m} t} \,  r^\ell \psi_{n \ell m}(r) Y_{\ell m} (\theta,\varphi) \, , \label{eq:single-wave}
\end{equation}

\noindent where $\gamma_{n\ell m}$ is a frequency to be determined from a well-posed eigenvalue problem. The quantum numbers that label each state take the values: $n=1,2,\ldots$, $\ell = 1, 2,\ldots,n-1$ and $m = -\ell,-\ell+1,\ldots,\ell-1,\ell$, where the number of nodes in the radial function $\psi_{n \ell m}$ is given by $n-1-\ell$. 

The gravitational potential is determined from the following Poisson equation,

\begin{equation}
    \nabla^2 V = \sum_{n \ell m} r^{2\ell} \psi^2_{n\ell m} Y_{\ell m} Y^\ast_{\ell m} \, . \label{eq:single-poisson}
\end{equation}

\noindent In order to solve Eq.~\eqref{eq:single-poisson}, it is convenient to consider an expansion of the gravitational potential in spherical harmonics of the form,

\begin{equation}
    V({\bf x}) = \sqrt{4\pi} \, \sum_{\ell m} V_{\ell m}(r) \, r^\ell  Y_{\ell m} (\theta,\varphi) \, . \label{eq:potential}
\end{equation}

\noindent Under the expansion~\eqref{eq:potential},  Poisson equation~\eqref{eq:single-poisson} becomes a set of equations for each radial function $V_{\ell m}$,

\begin{eqnarray}
    \nabla^2_{r\ell} V_{\ell 0} = \frac{\sqrt{4\pi}}{r^\ell} \sum_{n_1 \ell_1 m_1} (-1)^{m_1} \mathcal{G}^{\ell_1 \; \; \ell_1 \; \; \; \ell}_{m_1 \, -m_1 0} \, r^{2\ell_1} \psi^2_{n_1 \ell_1 m_1} \, . \label{eq:single-hierarchy}
\end{eqnarray}
where we have defined the $r\ell$-Laplacian operator $\nabla^2_{r \ell} = \partial^2_r + [2(\ell+1)/r] \partial_r$. 

Additionally, we have used in Eq.~\eqref{eq:potential} the so-called Gaunt coefficients $\mathcal{G}$, which are defined as \cite{Sebilleau1998}
\begin{equation}
    \mathcal{G}^{\ell_1 \; \ell_2 \; \; \ell}_{m_1 m_2 m} \equiv \int_\Omega Y_{\ell_1 m_1} Y_{\ell_2 m_2} Y^\ast_{\ell m} \, d\Omega \, . \label{eq:Gaunt}
\end{equation}

\noindent  Gaunt coefficients~\eqref{eq:Gaunt} follow the selection rules: $m=m_1+m_2$ and $|\ell_1 - \ell_2| \leq \ell \leq \ell_1 + \ell_2$, and are different from zero only if $\ell_1 + \ell_2 + \ell$ is an even number.

Notice that the magnetic number for all the radial coefficients in Eq.~\eqref{eq:single-hierarchy} is zero, which means that the gravitational potential does not depend on the azimuthal angle $\varphi$. This is a direct consequence of the selection rule on the magnetic number of the Gaunt coefficients~\eqref{eq:Gaunt}, which requires in this case that $m= m_1 -m_1 =0$. Additionally, the selection rules also read $0 \leq \ell \leq 2\ell_1$ and since the combination $2\ell_1 + \ell$ should be an even integer, then $\ell$ can only take even integer values: $\ell =0, 2, \ldots, 2\ell_1$.

On the other hand,  Schr\"odinger equation for the radial wave function $\psi_{n\ell m}$ in the ansatz (\ref{eq:single-wave}) is

\begin{equation}
    \nabla^2_{r\ell} \psi_{n \ell m} = 2 \left( \sqrt{4\pi} \sum_{\ell_1} \mathcal{G}^{\ell_1 \ell \; \ell}_{0 m m} \, r^{\ell_1} V_{\ell_1 0}  - \gamma_{n\ell m} \right) \psi_{n \ell m} \, . \label{eq:single-schr}
\end{equation}

\noindent A small note is in turn. To write down the foregoing equation we required the expansion of the product $V_{\ell_1 0} \Psi_{n \ell m}$ in terms of $Y_{\ell_2 m_2}$, which involves the Gaunt coefficients $\mathcal{G}^{\ell_1 \ell \; \ell_2}_{0 \, m m_2}$. Selection rules require  $m_2= m$ and $|\ell_1 - \ell| \leq \ell_2 \leq \ell_1 + \ell$. If $\ell_1 =0$, corresponding to the monopole term $V_{00}$, there is no other option but $\ell_2 = \ell$. However, if $\ell_1 \geq 2$, there is the possibility that $\ell_2$ can also take on larger values  than $\ell$, and then the expansion of the product $V_{\ell_1 0} \psi_{n \ell m}$ may have more non-zero terms than required for Eq.~\eqref{eq:single-schr}. In this respect, the latter should be considered an approximated expansion of Eq.~\eqref{eq:gpp} whenever $\ell_1 \geq 2$ (see also \cite{Silveira:1995dh} for a similar case).

To finish the description of our general framework, for the suggested ansatz~\eqref{eq:single-wave} we can calculate some physical quantities of interest. For instance, the total number of particles in the state $n \ell m$ is given by $N_{n \ell m} = (1/4\pi) \int |\Psi_{n \ell m}|^2 d^3 x$, whereas the kinetic and potential energies are respectively given by $K_{n\ell m} = -(1/2) \int \Psi^{\ast}_{n\ell m} \nabla^2 \Psi_{n\ell m} \, d^3x$ and $W_{n\ell m} = (1/2)\int |\Psi_{n\ell m}|^2 V \, d^3x$.

One can show from Eq.~\eqref{eq:single-schr} that for stationary configurations $K_{n \ell m} + 2W_{n \ell m} = \gamma_{n \ell m} N_{n \ell m}$, a relation that was first written down for spherically symmetric configurations \cite{Ruffini:1969qy,Seidel:1990jh,Guzman:2006yc}. Related quantities that will be useful below are the total energy $E_T = K_T + W_T$, and correspondingly the total kinetic and potential energies $K_T = \sum_{n\ell m} K_{n \ell m}$ and $W_T = \sum_{n\ell m} W_{n \ell m}$, respectively. Likewise, we define the (total) effective eigenfrequency as $\gamma_T N_T = \sum_{n\ell m} \gamma_{n \ell m} N_{n \ell m}$, where $N_T = \sum_{n\ell m} N_{n \ell m}$.

\begin{figure}[htp!]
\includegraphics[width=0.235\textwidth]{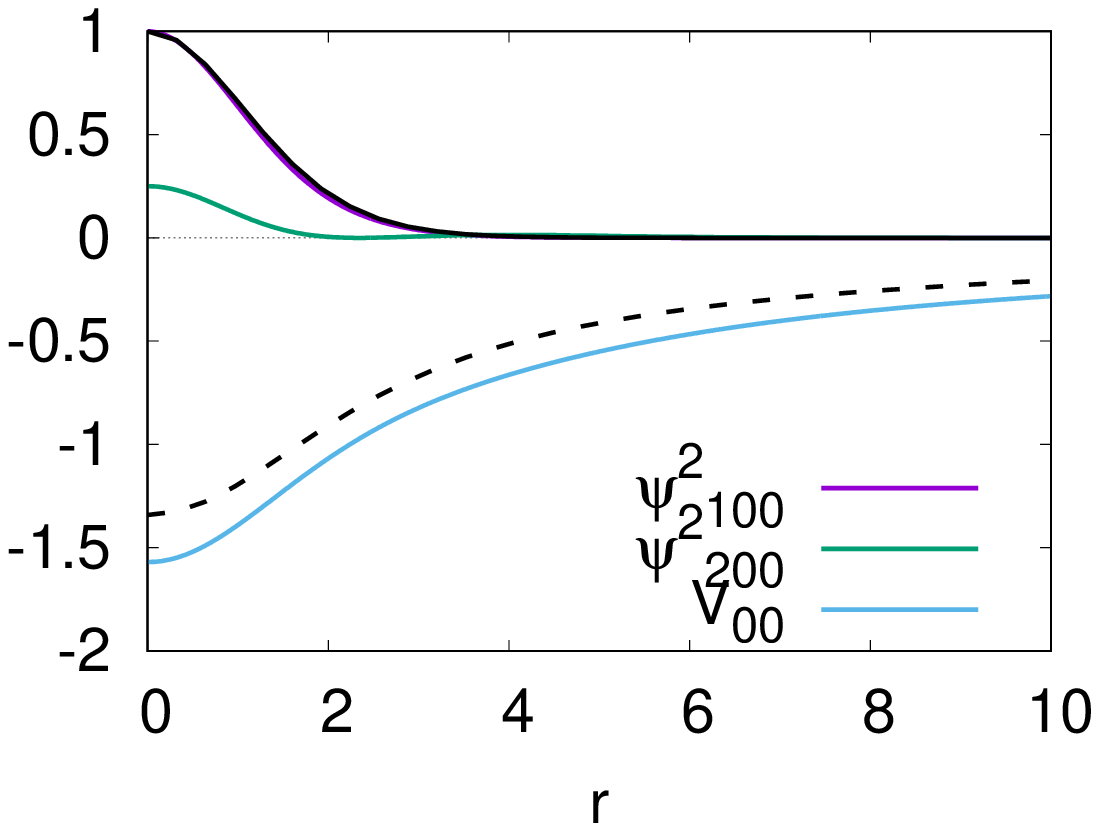}
\includegraphics[width=0.235\textwidth]{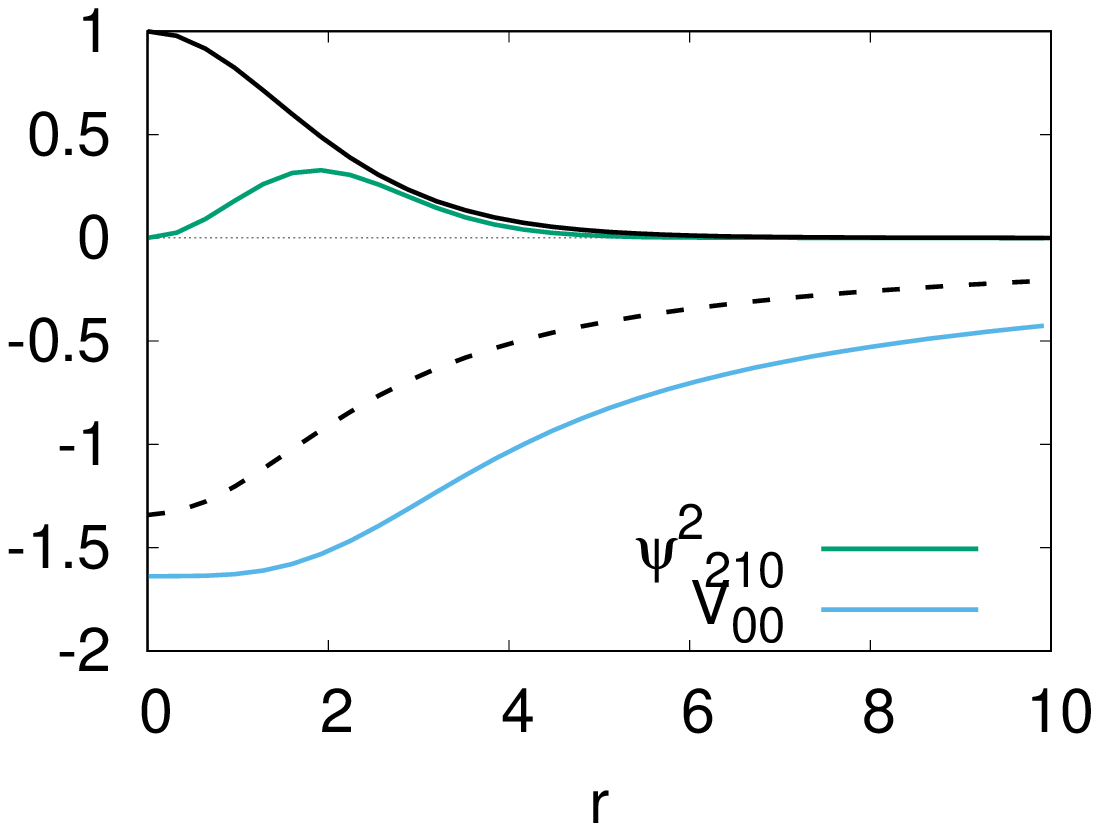}
\includegraphics[width=0.235\textwidth]{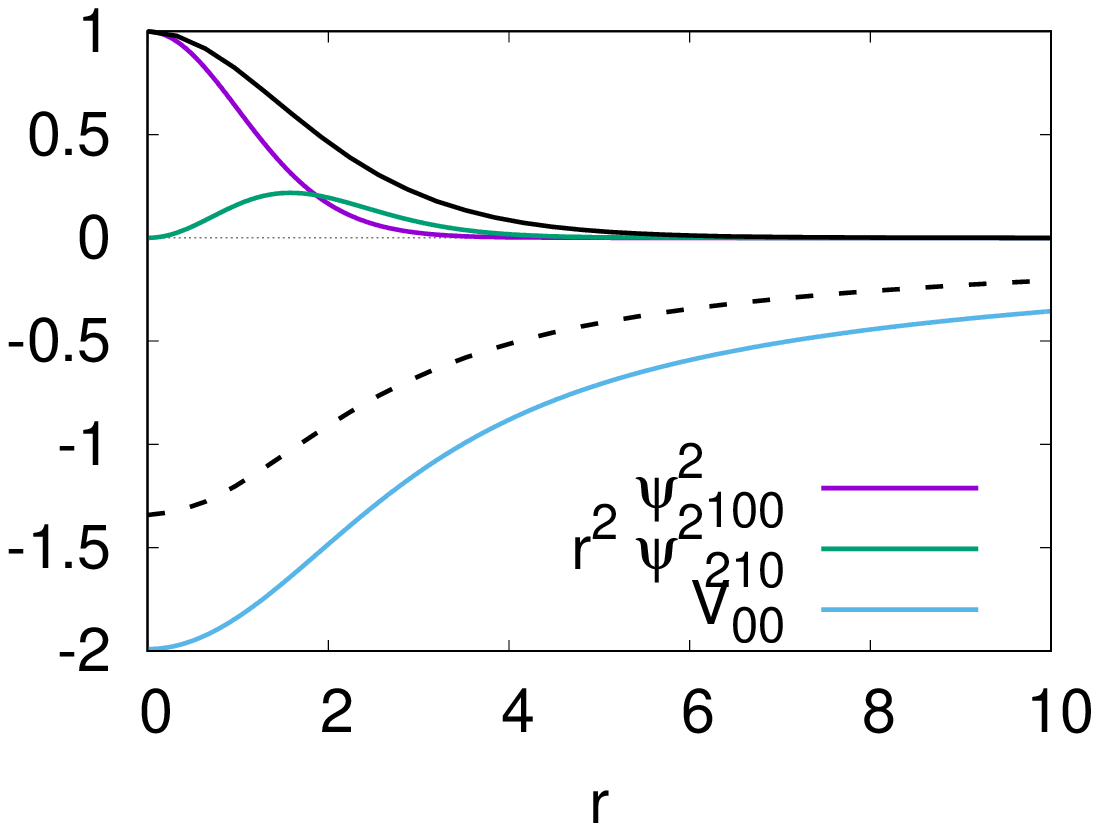}
\includegraphics[width=0.235\textwidth]{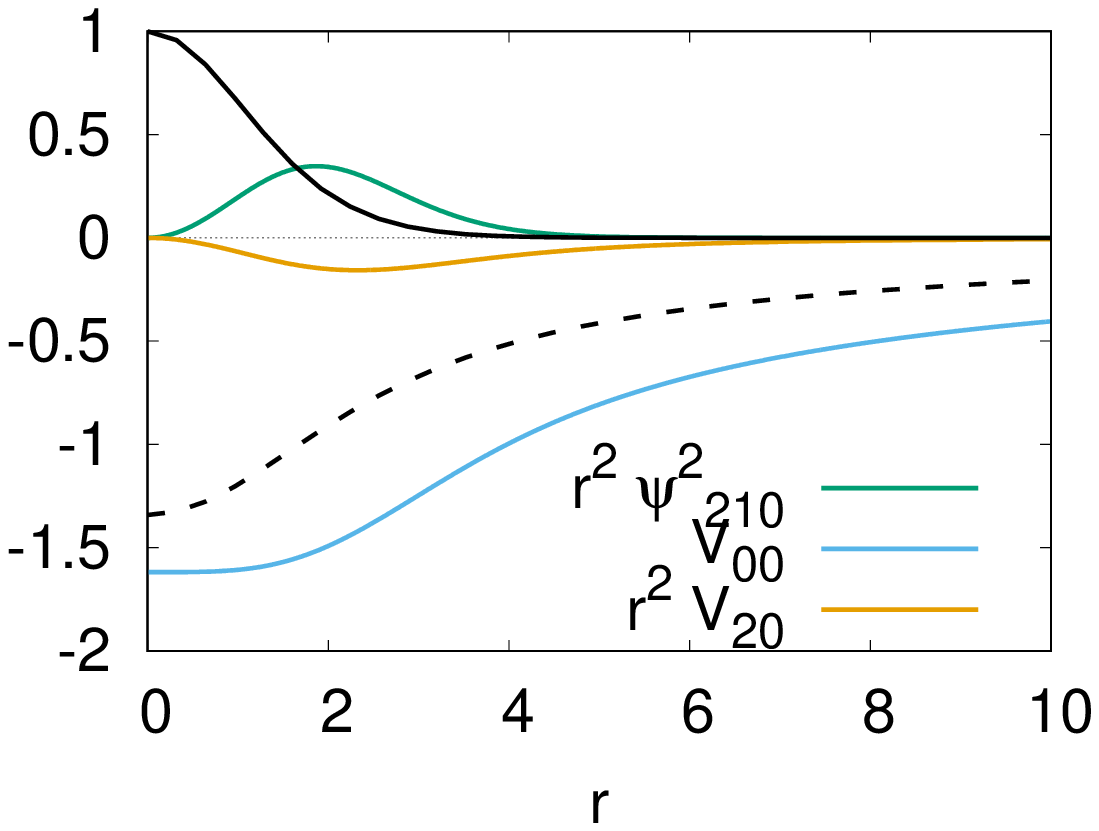}
\includegraphics[width=0.235\textwidth]{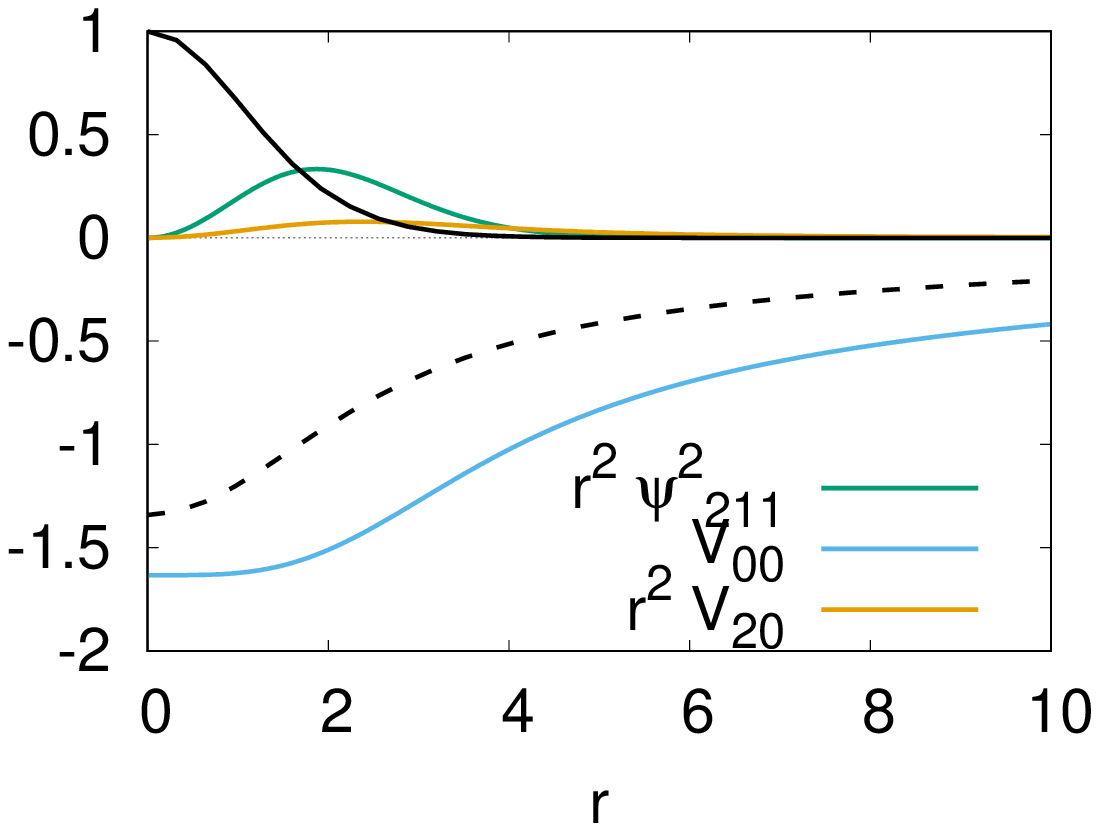}
\includegraphics[width=0.235\textwidth]{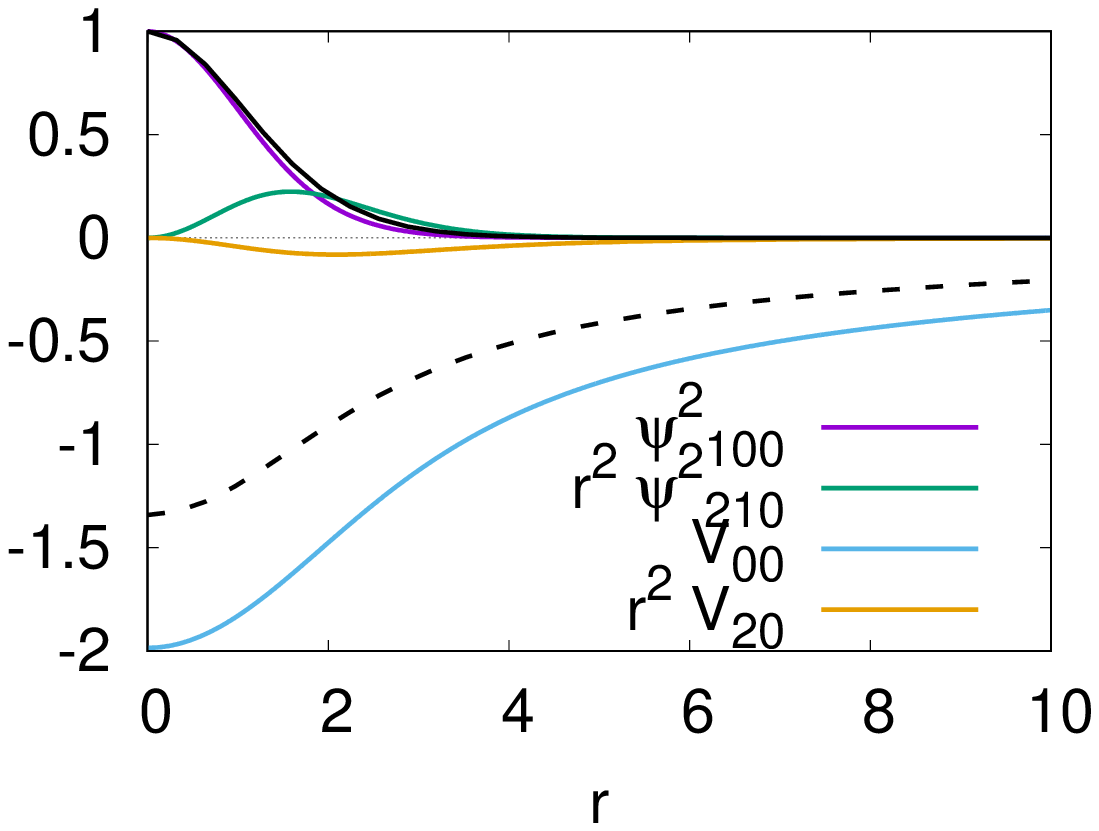}
    \caption{(Top Left) Double-state spherical configuration from Eqs.~\eqref{eq:multi-spherical}. Shown are the radial profiles  of $\psi^2_{100}$, $\psi_{200}$ and $V_{00}$. (Top Right) Dipole $\ell$-boson star from Eqs.~\eqref{eq:ell-star}, corresponding to the single state $\psi_{210}$. (Middle Left) Mixed $\ell$-boson star from Eqs.~\eqref{eq:ell-star}, corresponding to the states $\psi_{100}$ and $\psi_{210}$. In both instances of $\ell$-boson stars we also show the respective (monopole) gravitational potential $V_{00}$. (Middle Right) Dipole equilibrium configuration from Eqs.~\eqref{eq:dipole}. Shown are the radial profiles of $r^2 \psi^2_{210}$, $V_{00}$ and $r^2 V_{20}$. (Bottom Left) Dipole equilibrium configuration with angular momentum from Eqs.~\eqref{eq:dipole}. Shown are the radial profiles  of $r^2 \psi^2_{211}$, $V_{00}$ and $r^2 V_{20}$. (Bottom Right) Mixed state with a monopole-dipole combination from Eqs.~\eqref{eq:mixed}. Shown are the radial profiles of $\psi^2_{100}$, $\psi^2_{210}$, $V_{00}$ and $V_{20}$. For comparison in all cases, we show the radial profiles of $\psi^2_{100}$ and $V_{00}$ for a ground state equilibrium configuration, see Eqs.~\eqref{eq:spherical} (black solid and black dashed curves, respectively). See the text for more details.}
\label{fig:dipole}
\end{figure}

There are various interesting scenarios enclosed into this general framework that we are to describe now. We start first with a presentation of spherically symmetric cases, and then we continue with examples that  incorporate axially symmetric features. For all the cases we solve the equations of motion~\eqref{eq:single-hierarchy} and~\eqref{eq:single-schr} to find the equilibrium configurations and their particular properties. We have summarized different scenarios in Table~\ref{tab:general}, together with selected examples and the respective values of different quantities of interest. Likewise, we show in Fig.~\ref{fig:dipole} the radial profiles of wave functions and gravitational terms for some of the aforementioned examples.

\begin{table*}[htp]
\centering
\begin{tabular}{lccccccc}
\hline \hline
State $\psi_{n \ell m}$ & $\gamma_T$ & $N_T$ & $r_{95}$ & $K_T$ & $W_T$ & $K_T/|W_T|$ & Stability \\
\hline
1. Single spherical [Eqs.~\eqref{eq:spherical}] & & & & & & & \\
$\psi_{100}$ & $-0.69$ & $2.06$ & $3.93$ & $0.476$ & $-0.952$ & $1/2$ & Stable \\
$\psi_{200}$ & $-0.65$ & $4.59$ & $8.04$ & $0.990$ & $-1.981$ & $1/2$ & Unstable \\
$\cdots$ & & & & & & & Unstable \\
\hline
2. Multistate spherical [Eqs.~\eqref{eq:multi-spherical}] & & & & & & & \\
$(\psi_{100}, \psi_{200})$ & $-0.72$ & $2.82$ & $6.60$ & $0.681$ & $-1.353$ & $1/2$ & Stable if $N_{200}/N_{100} < 1.1$~\cite{UrenaLopez:2010ur} \\
\hline
3. $\ell$-boson star [Eqs.~\eqref{eq:ell-star}] & & & & & & & \\
$(\psi_{210},\psi_{211},\psi_{21-1})$ & $-0.96$ & $4.22$ & $4.80$ & $1.360$ & $-2.719$ & $1/2$ & Stable \\
$(\psi_{100},\psi_{210},\psi_{211},\psi_{21-1})$ & $-1.05$ & $3.55$ & $4.31$ & $1.239$ & $-2.478$ & $1/2$ & Stable \\
\hline
4. Single axial [Eqs.~\eqref{eq:dipole}] & & & & & & & \\
$\psi_{210}$ & $-1.04$ & $4.04$ & $4.70$ & $1.408$ & $-2.815$ & $1/2$ & Unstable \\
$\psi_{211} \, [\psi_{21-1}]$ & $-0.99$ & $4.18$ & $4.88$ & $1.373$ & $-2.745$ & $1/2$ & Unstable \\
\hline
5. Multistate [Eqs.~\eqref{eq:mixed}] & & & & & & \\
$(\psi_{100},\psi_{210})$ & $-1.07$ & $3.51$ & $4.20$ & $1.254$ & $-2.506$ & $1/2$ & Stable \\
$(\psi_{100},\psi_{211}) \, [(\psi_{100},\psi_{21-1})]$ & $-1.05$ & $3.54$ & $4.28$ & $1.243$ & $-2.485$ & $1/2$ & Stable \\
\hline \hline
\end{tabular}
\caption{\label{tab:general}Different physical quantities of the states described in the text. In order of appearance from left to right:  effective eigenfrequency $\gamma_T$,  total number of particles $N_T$,  $95\%$ radius $r_{95}$,  kinetic and potential energies $K_T$ and $W_T$, respectively,  and their virial ratio $K_T/|W_T|$. The values reported correspond to solutions of the  equations in each case, using the central values  $\psi_{100}(0)=1$ and $\psi_{210}(0) \, [\psi_{21\pm 1}(0)]=0.5$. The boundary conditions used at $r \to \infty$ were: $\psi_{n \ell m} = 0$, $V_{00}= -N_T/r$ and $V_{2 0}=0$, which in turn allowed the determination of the eigenvalues $\gamma_{n \ell m}$, $V_{00}(0)$ and $V_{20}(0)$. The stability was determined by numerically solving Eqs.~\eqref{eq:gpp}, verifying that the oscillation frequencies are consistent with the eigenvalue problem and that unitarity is preserved. We must recall that the superposition of the various wave functions assume the states they represent are independent.}
\end{table*}

\textit{1. Single state spherical configurations.} The first case is the single state solution with spherical symmetry, in which the wavefunction is $\psi_{n00}$, with $n \geq 1$. For the Gaunt coefficients in Poisson equation~\eqref{eq:single-hierarchy} we find $\mathcal{G}^{0 0 \ell}_{000} = \delta_{\ell 0}/\sqrt{4\pi}$, and then high multipole terms of the gravitational potential beyond the monopole  satisfy an homogeneous equation $\nabla^2_{r\ell} V_{\ell 0} =0$.  From this we can consider, without loss of generality, that $V_{\ell 0} =0$ for $\ell \geq 2$. Also, the only non-zero Gaunt coefficient for the Schr\"odinger equation~\eqref{eq:single-schr} is $\mathcal{G}^{0 0 0}_{000} = 1/\sqrt{4\pi}$. Thus, the equations of motion for spherically symmetric configurations are
\begin{eqnarray}
    \nabla^2_{r0} \psi_{n00} = 2 \left(V_{00} - \gamma_{n00} \right) \psi_{n00} \, , \quad 
    \nabla^2_{r0} V_{00} = \psi^2_{n00} \, . \label{eq:spherical}
\end{eqnarray}

Eqs.~\eqref{eq:spherical} conform a closed, self-contained system. Its solutions constitute a one parameter family of equilibrium configurations characterized by the central value of the wave function $\psi_{n00}(0)$. They have been widely studied and their stability properties are well established. Ground state equilibrium configurations ($n=1$) are stable, whereas excited configurations with nodes ($n\geq 2$) are unstable, even though all configurations are virialized, that is, the kinetic to potential energy ratio is  $K_{n00}/|W_{n00}|=1/2$ \cite{Ruffini:1969qy,Seidel:1990jh,Guzman:2004wj,Guzman:2006yc}.

\textit{2. Multistate spherical configurations.} 
In this case the equations of motion are

\begin{equation}
    \nabla^2_{r0} \psi_{n00} = 2 \left(V_{00} - \gamma_{n00} \right) \psi_{n00} \, , \; 
    \nabla^2_{r0} V_{00} = \sum_n \psi^2_{n00} \, . \label{eq:multi-spherical}
\end{equation}
These multistate configurations conform a multi-parameter family of solutions characterised by the central value of each wave function: $\psi_{100}(0),\psi_{200}(0),\ldots$ It was found in Ref.~\cite{UrenaLopez:2010ur} that in the case of two state configurations, stablility is granted for $N_{100}/N_{200}<1.1$. Although the states are not virialized separately, that is $2K_{n00}+W_{n00} \neq 0$, the total kinetic and potential energies  satisfy the virial relation $\sum_n (2K_{n00}+W_{n00})= 2K_T + W_T= 0$. This means that, collectively, multistate configurations also satisfy the energy relations found for single configurations, in particular that their total energy is related to the potential and number of particles in the form: $E_T = (1/2) W_T = (1/3) \gamma_T N_T$.


{\it 3. Non-relativistic $\ell$-boson stars.} 
Our approach includes also the so-called $\ell$-boson stars \cite{Olabarrieta:2007di,Alcubierre:2018ahf,Alcubierre:2019qnh}  in the Newtonian limit. For this case, the radial functions are the same for all possible values of the magnetic number, that is, $\psi_{n\ell m} = \psi_{n \ell 0}$. The hierarchy of equations~\eqref{eq:single-hierarchy} under this particular assumption reads

\begin{subequations}
\label{eq:ell-star}
\begin{eqnarray}
        \nabla^2_{r\ell} V_{\ell 0} &=& \frac{\sqrt{4\pi}}{r^\ell} \sum_{n_1,\ell_1} r^{2\ell_1} \psi^2_{n_1 \ell_1 0} \left[ \sum^{\ell_1}_{m_1 = -\ell_1} (-1)^{m_1} \mathcal{G}^{\ell_1 \; \; \; \ell_1 \; \; \; \ell}_{m_1 -m_1 0} \right] \nonumber \\
        &=& \frac{\delta_{\ell 0}}{r^\ell} \sum_{n_1,\ell_1} (2\ell_1 +1) \, r^{2\ell_1} \psi^2_{n_1 \ell_1 0} \, . \label{eq:ell-star-a}
\end{eqnarray}
where we have used the standard addition theorem of spherical harmonics. The Kronecker delta in Eq.~\eqref{eq:ell-star-a} implies that the only surviving multipole term of the gravitational potential is the monopolar one, $V_{00}$. The complete set of equations in this case is complemented by the hierarchy of Schr\"odinger equations in the form
\begin{equation}
    \nabla^2_{r\ell} \psi_{n \ell 0} = 2 \left( V_{00} - \gamma_{n\ell 0} \right) \psi_{n \ell 0} \, . \label{eq:ell-star-b}
\end{equation}
\end{subequations}

Eqs.~\eqref{eq:ell-star} confirm, first, that the Newtonian gravitational potential of $\ell$-boson stars is spherically symmetric, just as their relativistic counterparts \cite{Olabarrieta:2007di,Alcubierre:2018ahf,Alcubierre:2019qnh}; and, second, that one can also consider multistate $\ell$-boson stars. Notice the resemblance of Eqs.~\eqref{eq:ell-star} with Eqs.~\eqref{eq:multi-spherical}: multistate $\ell$-boson stars become a generalization of the multistate spherical configurations, but now with the involvement of axially-symmetric density profiles.

{\it 4. Single axially-symmetric configurations}. These  are a different generalization from single state spherical solutions. We illustrate the solution with the single dipolar $\psi_{210} Y_{10}$ term in Eq. (\ref{eq:single-wave}). For the right hand side of  Poisson equation~\eqref{eq:single-hierarchy} we require the Gaunt coefficients $\mathcal{G}^{110}_{000}=1/\sqrt{4\pi}$ and $\mathcal{G}^{112}_{000}=1/(\sqrt{5\pi})$, which implies that the gravitational potential must be represented by the monopolar  $V_{00}$ and  quadrupololar  $V_{20}$ terms only. For  Schr\"odinger equation~\eqref{eq:single-schr} we need the Gaunt coefficients $\mathcal{G}^{011}_{000}=1/\sqrt{4\pi}$ and $\mathcal{G}^{211}_{000}=1/(\sqrt{5\pi})$. Thus, the SP system splits into the following system of equations\footnote{Following the small note after Eq.~\eqref{eq:single-schr}, another non-zero Gaunt coefficient that arises in the expansion of Eq.~\eqref{eq:dipole-a} is $\mathcal{G}^{2 1 3}_{0 0 0} = (3/2)\sqrt{3/35\pi}$. This coefficient implies the presence of a term of the form $V_{20} \psi_{210} Y_{30}$ that could not be included in Eq.~\eqref{eq:dipole-a}, and then the latter must be seen as an approximated representation of Eq.~\eqref{eq:gpp} for the dipole wavefunction $\Psi_{210}$.},

\begin{subequations}
\label{eq:dipole}
\begin{eqnarray}
    \nabla^2_{r1} \psi_{210} &=& 2 \left( V_{00} + \frac{2}{\sqrt{5}} r^2 V_{20} - \gamma_{210} \right) \psi_{210} \, , \label{eq:dipole-a} \\
    \nabla^2_{r0} V_{00} &=& r^2 \psi^2_{210} \, , \quad \nabla^2_{r2} V_{20} = \frac{2}{\sqrt{5}} \psi^2_{210} \, . \label{eq:dipole-b}
\end{eqnarray}
\end{subequations}

We can include angular momentum by considering  the single wavefunction $\psi_{211}$. The required Gaunt coefficients now are: $\mathcal{G}^{1 \;  \; 10}_{1-10}=-1/\sqrt{4\pi}$ and $\mathcal{G}^{1 \; \; 12}_{1-10}=1/(2\sqrt{5\pi})$ for  Poisson equation~\eqref{eq:single-hierarchy}; and $\mathcal{G}^{011}_{011}=1/\sqrt{4\pi}$ and $\mathcal{G}^{211}_{011}=-1/(2\sqrt{5\pi})$ for Schr\"odinger equation~\eqref{eq:single-schr}. Hence, the resulting equations of motion for a rotating dipole are obtained from Eqs.~\eqref{eq:dipole} by the mere replacements $\psi_{210} \to \psi_{211}$, and $2/\sqrt{5} \to -1/\sqrt{5}$ for the terms involving $V_{20}$. The change of sign means that the inclusion of angular momentum in the dipole configuration has the effect to make the (quadrupole) gravitational potential $V_{20}$ repulsive.\footnote{The inclusion of angular momentum may allow the formation of vortices in equilibrium configurations, which in turn can be useful in studies of dark matter with Bose-Einstein condensates, see for instance~\cite{2012MNRAS.422..135R}.} 

\textit{5. Multistate axial configurations.} We are in position to construct configurations with arbitrary combinations of wave functions, either spherically or axially symmetric. As a representative example, we consider a mixed configuration composed of monopole and dipole components. Taking into account the previously calculated Gaunt coefficients, Eqs.~\eqref{eq:single-hierarchy} and~\eqref{eq:single-schr} become now four equations,

\begin{subequations}
\label{eq:mixed}
\begin{eqnarray}
    \nabla^2_{r0} \psi_{100} &=& 2 \left( V_{00}- \gamma_{100} \right) \psi_{100} \, , \\
    \nabla^2_{r1} \psi_{210} &=& 2 \left( V_{00} + \frac{2}{\sqrt{5}} r^2 V_{20} - \gamma_{210} \right) \psi_{210} \, , \\
    \nabla^2_{r0} V_{00} &=& \psi^2_{100} + r^2 \psi^2_{210} \, , \quad     \nabla^2_{r2} V_{20} = \frac{2}{\sqrt{5}} \psi^2_{210} \, .
\end{eqnarray}
\end{subequations}
If we were to consider the mixed state with angular momentum, e.g. $\psi_{100}$ and $\psi_{211}$, we only need to replace $2/\sqrt{5} \to -1/\sqrt{5}$ for the terms involving $V_{20}$ in Eqs.~\eqref{eq:mixed}. The solution is parametrized by the central values $\psi_{100}(0)$ and $\psi_{210}(0)$ and in general on the central value of the wave functions associated to each state.

\textit{Stability.} In order to check the stability of the configurations we solve the full time-dependent system (\ref{eq:gpp}) for the various configurations, using an enhanced version of the 3D code in Ref.~\cite{Guzman:2013rua} that evolves now multiple states. Should a configuration be long-lived with diagnostics evolving around equilibrium values is our criteria to determine stability/instability of the cases shown in Table~\ref{tab:general}. 

\begin{figure}[htp!]
\includegraphics[width=4.2cm]{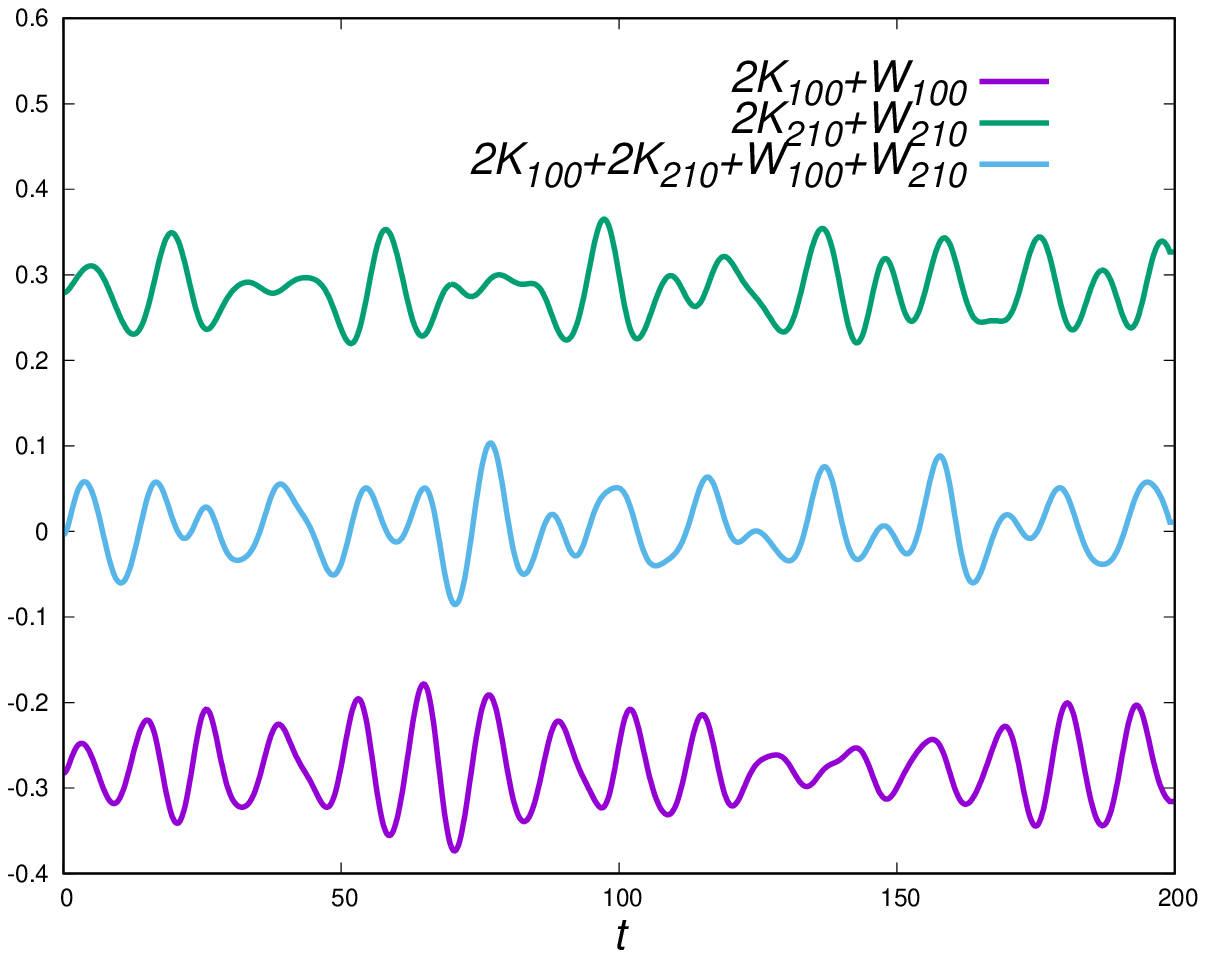}
\includegraphics[width=4.2cm]{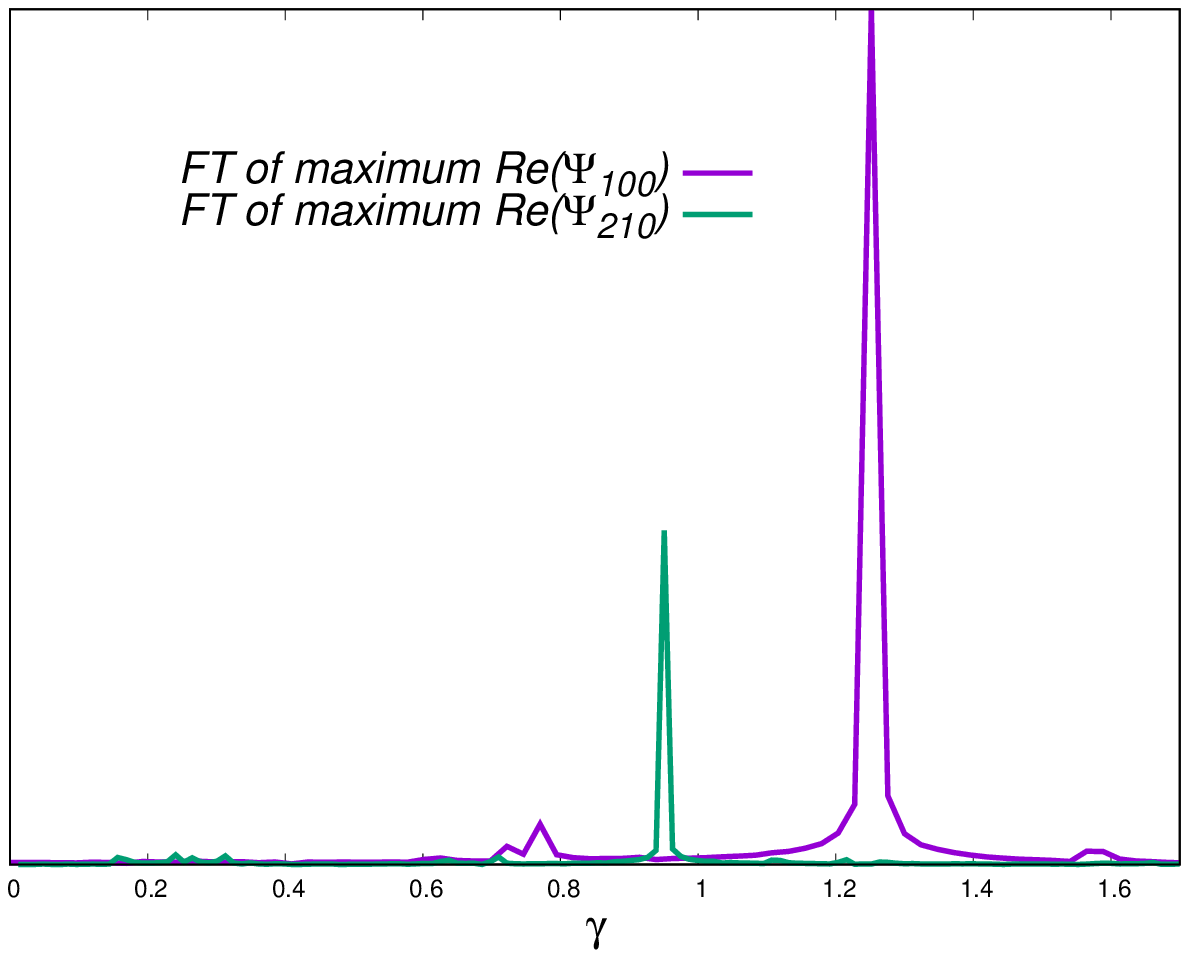}
\includegraphics[width=4.2cm]{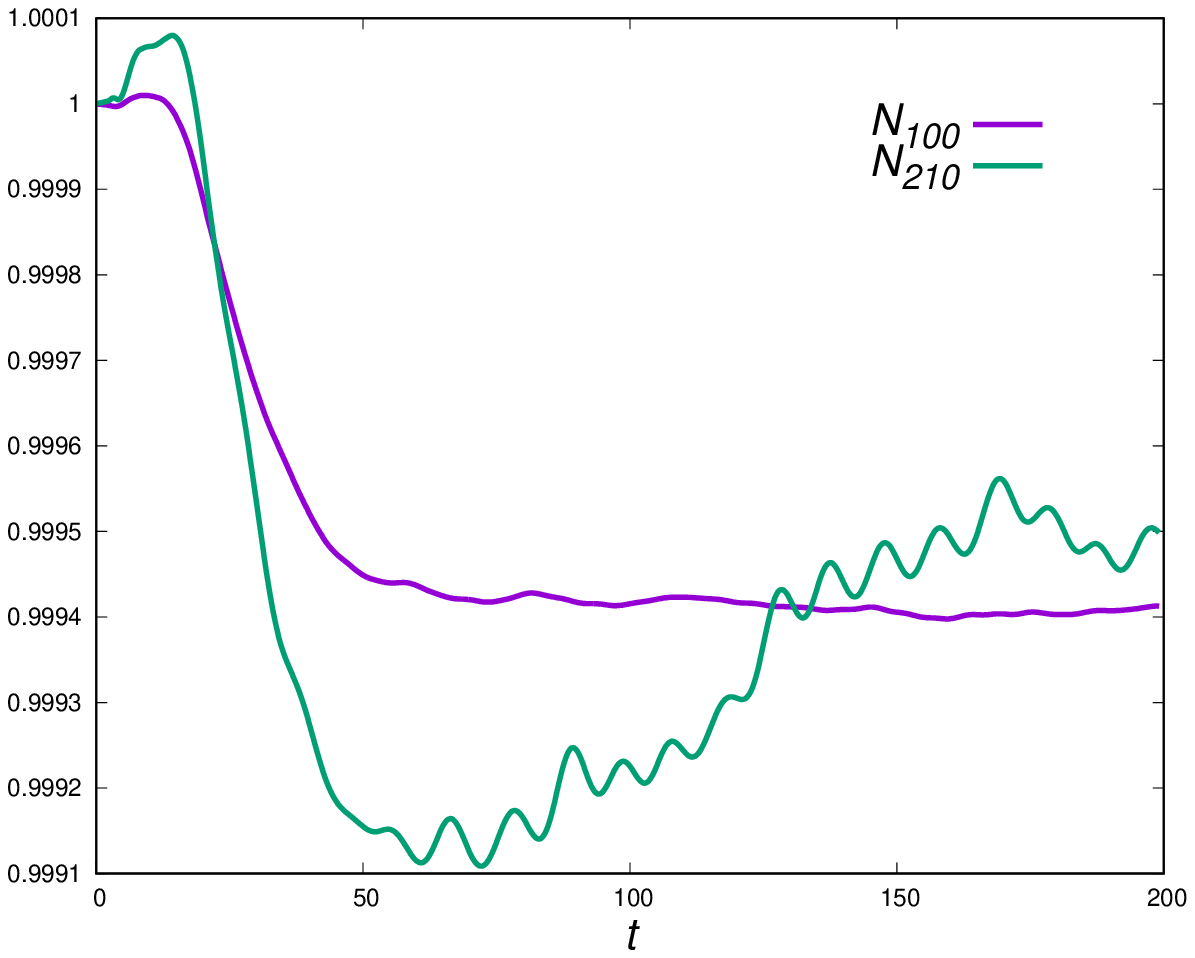}
\includegraphics[width=4.2cm]{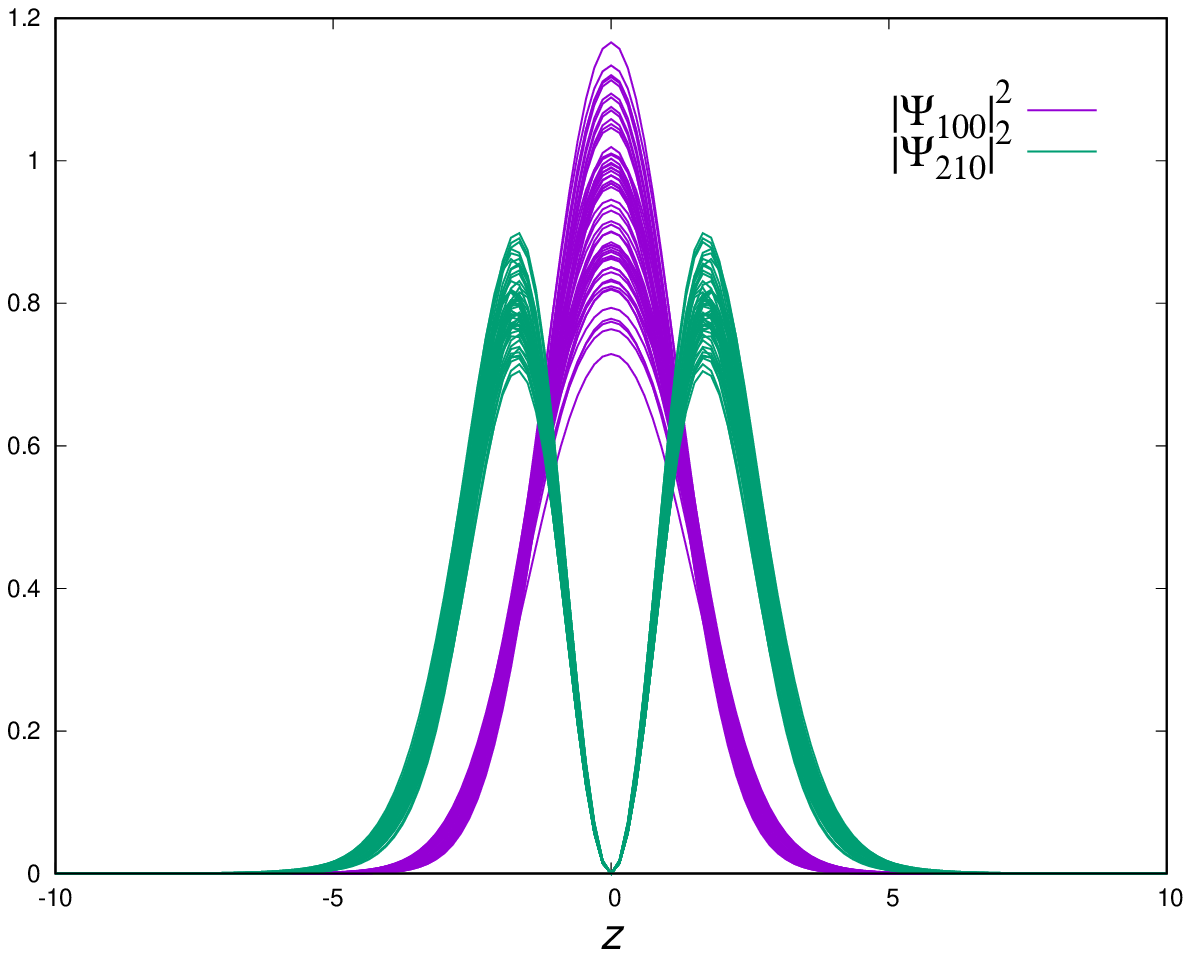}
\caption{Diagnostics of the mixed monopole-dipole configuration. In the top left panel we show separately the quantities $2K_{100}+W_{100}$ and $2K_{210}+W_{210}$, so as their addition. In the top right panel we show the Fourier Transform of the central value of $\Psi_{100}$ and the maximum of $\Psi_{210}$ as functions of time. Then we show in the bottom left the number of particles in each state varies by less than 0.1\%, and finally snapshots of the density of each state in the bottom right panel that show how they are genuinely evolving.}
\label{fig:mixed_diagnostics}
\end{figure}

As a representative case, we show the evolution of a two-state axial configuration of the form $\psi_{100}+\psi_{210}$ in Fig.~\ref{fig:mixed_diagnostics}. The time window used for the evolution is $t \in [0,300]$ which includes about sixty cycles of the spherical wavefunction $\psi_{100}$. Two important quantities of the evolution are shown in the top left panel of Fig. \ref{fig:mixed_diagnostics}, which correspond to the energy combinations $2K_{100}+W_{100}$ and $2K_{210}+W_{210}$. They are not zero, but the total quantity $2K_T+W_T$ oscillates around zero as expected for (nearly) virialized systems. Another important diagnostics consists in verifying that the wavefunctions oscillate with their expected eigenfrequencies. For this calculate the Fourier Transform of the maximum values of $\psi_{100}$ and $\psi_{210}$ as functions of time. We see from the right top panel in Fig.~\ref{fig:mixed_diagnostics} that the eigenfrequencies are $\gamma_{100}\simeq 1.25$ and $\gamma_{210}\simeq 0.93$, whereas we measure  $N_{100}=1.537$ and $N_{210}=1.9714$. These results together imply that the effective frequency obtained from the evolution $\gamma_T=\frac{\gamma_{100}N_{100}+\gamma_{210}N_{210}}{N_{100}+N_{210}}=1.07$, is in agreement with the solution of the eigenvalue problem at initial time reported in Table~\ref{tab:general}.

We also checked unitarity through the conservation of the number of particles in each state, namely $N_{100}$ and $N_{210}$. It can be seen from the bottom left panel in Fig.~\ref{fig:mixed_diagnostics} that the  number of particles in each state, normalized to their initial value, changes in less than 0.1\%. Finally, a sign of evolution is that the density of the two states oscillate conspiring to maintain the configuration long-lived, as shown in the bottom right panel in Fig.~\ref{fig:mixed_diagnostics}.

{\it Formation of multistate configurations.} If these configurations are to play a role in astrophysics and cosmology, it is important to show not only that they are long-living solutions, but that they can be formed. For this we think of a simple scenario where two equilibrium configurations merge, with the condition that they are made of two different non-coherent fields $\Psi^{(1)}_{100}$ and $\Psi^{(2)}_{100}$, associated to equilibrium spherical solutions in the ground state. Consistently with (\ref{eq:gpp}), we assume the system evolves according to $i\dot{\Psi}^{(j)}_{100} = -\frac{1}{2}\nabla^2 \Psi^{(j)}_{100} + V\Psi^{(j)}_{100}$ with $j=1,2$ and $\nabla^2 V = \sum_{j=1}^{2}|\Psi^{(j)}_{100} |^2$. Snapshots of an unequal mass head-on merger are shown in Fig.~\ref{fig:merger} and animations appear in the supplemental material \cite{supplemental}. After the encounter of the two configurations, the smaller configuration splits into two regions along the head-on axis $z$. The system oscillates around the center of mass during a time window corresponding to more than 2000 cycles of the wave function. During the evolution the system does not settle into a nearly stationary configuration, however the morphology of the densities, even if time-dependent,  resembles that in the bottom right panel of Fig.~\ref{fig:mixed_diagnostics}.

\begin{figure}
\includegraphics[width=0.235\textwidth]{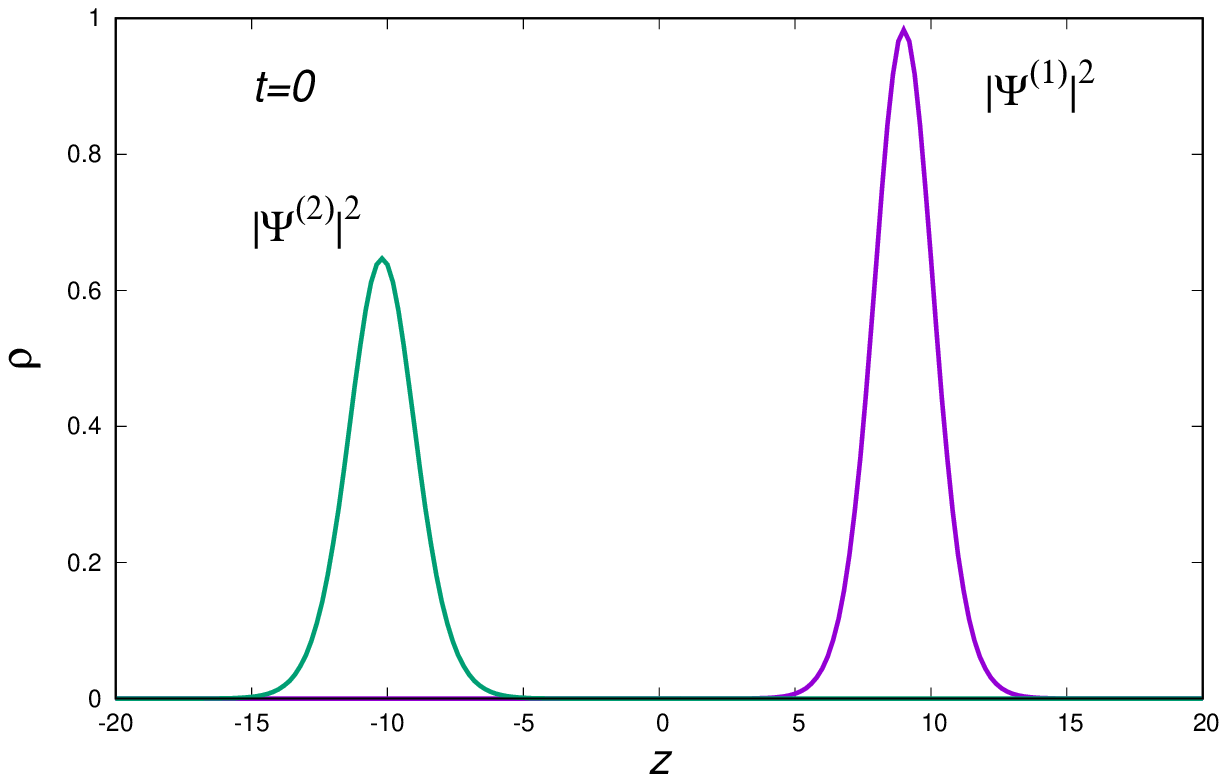}
\includegraphics[width=0.235\textwidth]{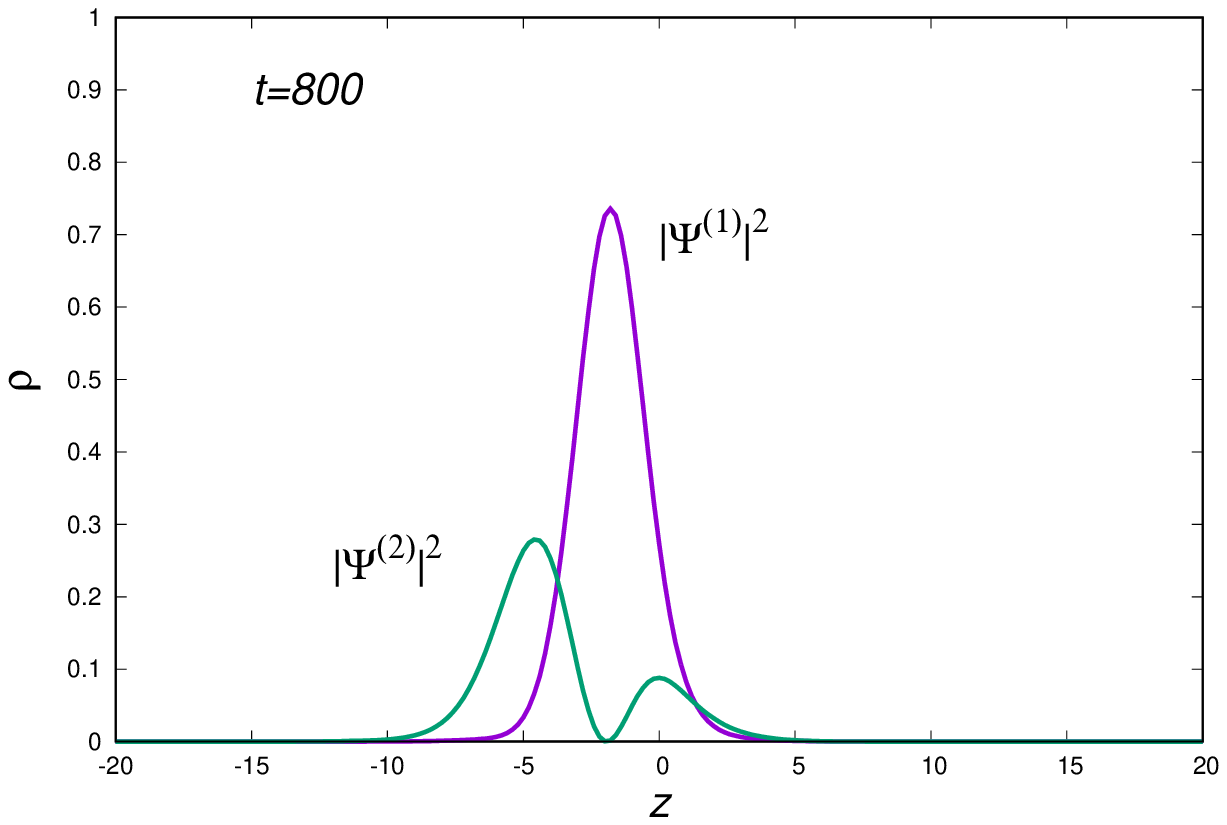}
\includegraphics[width=0.235\textwidth]{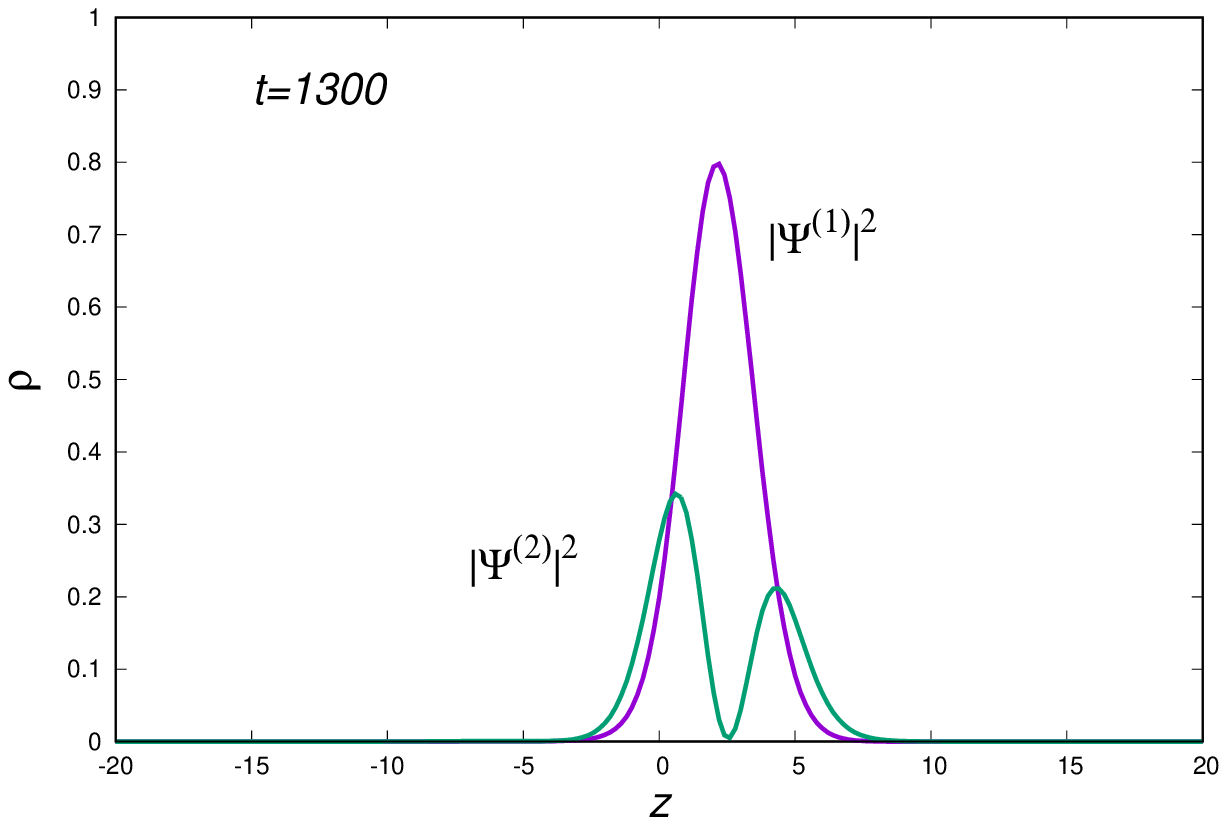}
\includegraphics[width=0.235\textwidth]{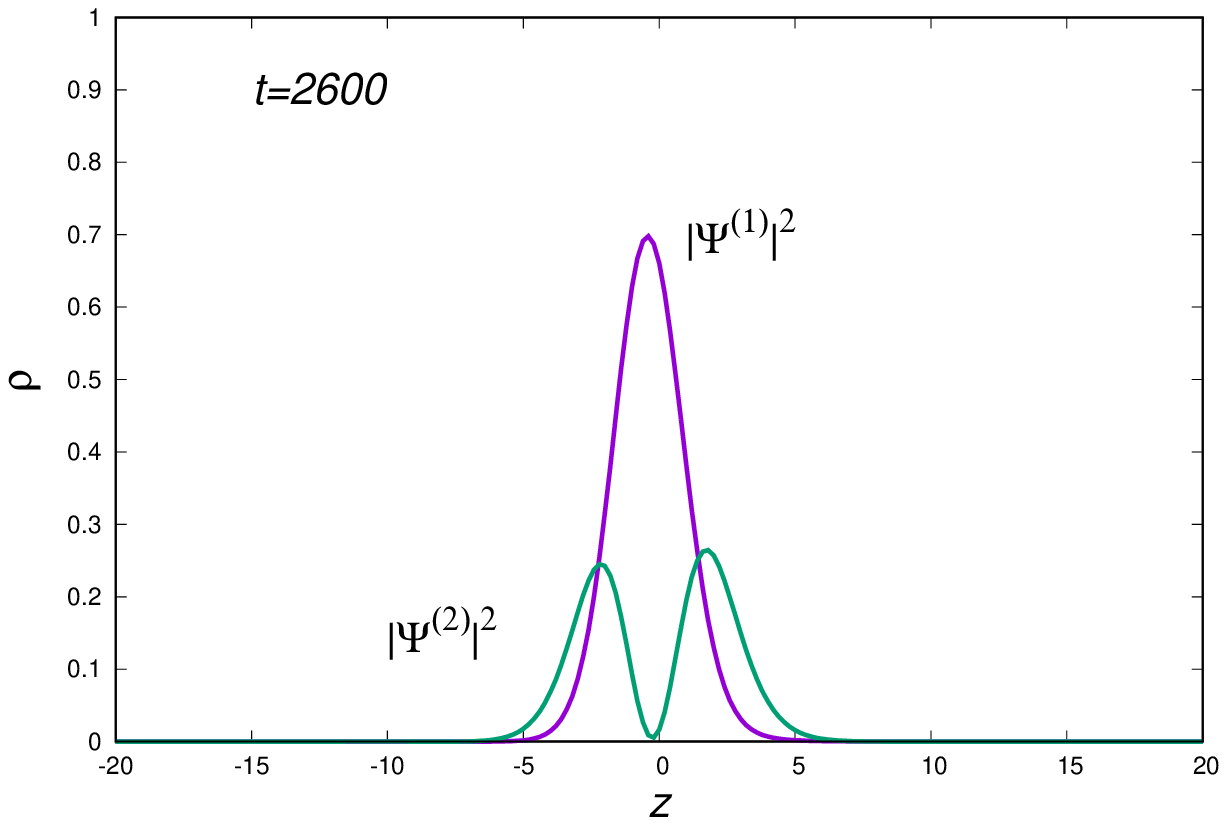}
\caption{Snapshots of the merger of two spherical equilibrium configurations in different coherent states that share the gravitational potential. In this case we show the case of mass-ratio $N^{(1)}/N^{(2)}=0.9$ and initial head-on momentum along the $z$-axis of $p_z=0.3$, using the conventions in Ref.~\cite{PhysRevD.97.116003}.} \label{fig:merger}
\end{figure}


{\it Final comments.} We presented a general approach for the construction of axially symmetric multistate solutions of the SP system, with and without angular momentum. We also sampled the properties of particular representative configurations of single and two-state configurations for illustration, whose stability was studied based on numerical evolutions. We expect this method to have impact on studies related to anisotropic halos of bosonic dark matter, since they might provide an explanation of dwarf galaxy distributions \cite{Satellites}, reason why a  formation process of multistate configurations would be useful.


\section*{Acknowledgments}
LAU-L was partially supported by Programa para el Desarrollo Profesional Docente; Direcci\'on de Apoyo a la Investigaci\'on y al Posgrado, Universidad de Guanajuato; CONACyT M\'exico under Grant No. A1-S-17899; and the Instituto Avanzado de Cosmolog\'ia Collaboration. This research is supported by grants CIC-UMSNH-4.9, CONACyT: 258726, A1-S-8742, 269652 and CB-2014-1, No. 240512. The runs were carried out in the Big Mamma cluster at the Laboratorio de Inteligencia Artificial y Superc\'omputo, IFM-UMSNH. 


\bibliography{Paper_final}

\begin{thebibliography}{36}%
\makeatletter
\providecommand \@ifxundefined [1]{%
 \@ifx{#1\undefined}
}%
\providecommand \@ifnum [1]{%
 \ifnum #1\expandafter \@firstoftwo
 \else \expandafter \@secondoftwo
 \fi
}%
\providecommand \@ifx [1]{%
 \ifx #1\expandafter \@firstoftwo
 \else \expandafter \@secondoftwo
 \fi
}%
\providecommand \natexlab [1]{#1}%
\providecommand \enquote  [1]{``#1''}%
\providecommand \bibnamefont  [1]{#1}%
\providecommand \bibfnamefont [1]{#1}%
\providecommand \citenamefont [1]{#1}%
\providecommand \href@noop [0]{\@secondoftwo}%
\providecommand \href [0]{\begingroup \@sanitize@url \@href}%
\providecommand \@href[1]{\@@startlink{#1}\@@href}%
\providecommand \@@href[1]{\endgroup#1\@@endlink}%
\providecommand \@sanitize@url [0]{\catcode `\\12\catcode `\$12\catcode
  `\&12\catcode `\#12\catcode `\^12\catcode `\_12\catcode `\%12\relax}%
\providecommand \@@startlink[1]{}%
\providecommand \@@endlink[0]{}%
\providecommand \url  [0]{\begingroup\@sanitize@url \@url }%
\providecommand \@url [1]{\endgroup\@href {#1}{\urlprefix }}%
\providecommand \urlprefix  [0]{URL }%
\providecommand \Eprint [0]{\href }%
\providecommand \doibase [0]{http://dx.doi.org/}%
\providecommand \selectlanguage [0]{\@gobble}%
\providecommand \bibinfo  [0]{\@secondoftwo}%
\providecommand \bibfield  [0]{\@secondoftwo}%
\providecommand \translation [1]{[#1]}%
\providecommand \BibitemOpen [0]{}%
\providecommand \bibitemStop [0]{}%
\providecommand \bibitemNoStop [0]{.\EOS\space}%
\providecommand \EOS [0]{\spacefactor3000\relax}%
\providecommand \BibitemShut  [1]{\csname bibitem#1\endcsname}%
\let\auto@bib@innerbib\@empty
\bibitem [{\citenamefont {Ruffini}\ and\ \citenamefont
  {Bonazzola}(1969)}]{Ruffini:1969qy}%
  \BibitemOpen
  \bibfield  {author} {\bibinfo {author} {\bibfnamefont {R.}~\bibnamefont
  {Ruffini}}\ and\ \bibinfo {author} {\bibfnamefont {S.}~\bibnamefont
  {Bonazzola}},\ }\href {\doibase 10.1103/PhysRev.187.1767} {\bibfield
  {journal} {\bibinfo  {journal} {Phys. Rev.}\ }\textbf {\bibinfo {volume}
  {187}},\ \bibinfo {pages} {1767} (\bibinfo {year} {1969})}\BibitemShut
  {NoStop}%
\bibitem [{\citenamefont {Schunck}\ and\ \citenamefont
  {Mielke}(2003)}]{Schunck:2003kk}%
  \BibitemOpen
  \bibfield  {author} {\bibinfo {author} {\bibfnamefont {F.~E.}\ \bibnamefont
  {Schunck}}\ and\ \bibinfo {author} {\bibfnamefont {E.~W.}\ \bibnamefont
  {Mielke}},\ }\href {\doibase 10.1088/0264-9381/20/20/201} {\bibfield
  {journal} {\bibinfo  {journal} {Class. Quant. Grav.}\ }\textbf {\bibinfo
  {volume} {20}},\ \bibinfo {pages} {R301} (\bibinfo {year} {2003})},\ \Eprint
  {http://arxiv.org/abs/0801.0307} {arXiv:0801.0307 [astro-ph]} \BibitemShut
  {NoStop}%
\bibitem [{\citenamefont {Liebling}\ and\ \citenamefont
  {Palenzuela}(2012)}]{Liebling:2012fv}%
  \BibitemOpen
  \bibfield  {author} {\bibinfo {author} {\bibfnamefont {S.~L.}\ \bibnamefont
  {Liebling}}\ and\ \bibinfo {author} {\bibfnamefont {C.}~\bibnamefont
  {Palenzuela}},\ }\href {\doibase 10.12942/lrr-2012-6,
  10.1007/s41114-017-0007-y} {\bibfield  {journal} {\bibinfo  {journal} {Living
  Rev. Rel.}\ }\textbf {\bibinfo {volume} {15}},\ \bibinfo {pages} {6}
  (\bibinfo {year} {2012})},\ \bibinfo {note} {[Living Rev.
  Rel.20,no.1,5(2017)]},\ \Eprint {http://arxiv.org/abs/1202.5809}
  {arXiv:1202.5809 [gr-qc]} \BibitemShut {NoStop}%
\bibitem [{\citenamefont {Mielke}(2016)}]{Mielke:2016war}%
  \BibitemOpen
  \bibfield  {author} {\bibinfo {author} {\bibfnamefont {E.~W.}\ \bibnamefont
  {Mielke}},\ }\href {\doibase 10.1007/978-3-319-31299-6_6} {\bibfield
  {journal} {\bibinfo  {journal} {Fundam. Theor. Phys.}\ }\textbf {\bibinfo
  {volume} {183}},\ \bibinfo {pages} {115} (\bibinfo {year}
  {2016})}\BibitemShut {NoStop}%
\bibitem [{\citenamefont {{Schupp}}\ and\ \citenamefont {{van der
  Bij}}(1996)}]{1996PhLB..366...85S}%
  \BibitemOpen
  \bibfield  {author} {\bibinfo {author} {\bibfnamefont {B.}~\bibnamefont
  {{Schupp}}}\ and\ \bibinfo {author} {\bibfnamefont {J.~J.}\ \bibnamefont
  {{van der Bij}}},\ }\href {\doibase 10.1016/0370-2693(95)01327-X} {\bibfield
  {journal} {\bibinfo  {journal} {Physics Letters B}\ }\textbf {\bibinfo
  {volume} {366}},\ \bibinfo {pages} {85} (\bibinfo {year} {1996})},\ \Eprint
  {http://arxiv.org/abs/astro-ph/9508017} {arXiv:astro-ph/9508017 [astro-ph]}
  \BibitemShut {NoStop}%
\bibitem [{\citenamefont {Matos}\ \emph {et~al.}(2009)\citenamefont {Matos},
  \citenamefont {Vazquez-Gonzalez},\ and\ \citenamefont
  {Magana}}]{Matos:2008ag}%
  \BibitemOpen
  \bibfield  {author} {\bibinfo {author} {\bibfnamefont {T.}~\bibnamefont
  {Matos}}, \bibinfo {author} {\bibfnamefont {A.}~\bibnamefont
  {Vazquez-Gonzalez}}, \ and\ \bibinfo {author} {\bibfnamefont
  {J.}~\bibnamefont {Magana}},\ }\href {\doibase
  10.1111/j.1365-2966.2008.13957.x} {\bibfield  {journal} {\bibinfo  {journal}
  {Mon. Not. Roy. Astron. Soc.}\ }\textbf {\bibinfo {volume} {393}},\ \bibinfo
  {pages} {1359} (\bibinfo {year} {2009})},\ \Eprint
  {http://arxiv.org/abs/0806.0683} {arXiv:0806.0683 [astro-ph]} \BibitemShut
  {NoStop}%
\bibitem [{\citenamefont {Suarez}\ \emph {et~al.}(2014)\citenamefont {Suarez},
  \citenamefont {Robles},\ and\ \citenamefont {Matos}}]{Suarez:2013iw}%
  \BibitemOpen
  \bibfield  {author} {\bibinfo {author} {\bibfnamefont {A.}~\bibnamefont
  {Suarez}}, \bibinfo {author} {\bibfnamefont {V.~H.}\ \bibnamefont {Robles}},
  \ and\ \bibinfo {author} {\bibfnamefont {T.}~\bibnamefont {Matos}},\
  }\bibfield  {booktitle} {\emph {\bibinfo {booktitle} {{Proceedings of 4th
  International Meeting on Gravitation and Cosmology (MGC 4): Santa Clara,
  Cuba, June 1-4, 2009}}},\ }\href {\doibase 10.1007/978-3-319-02063-1_9}
  {\bibfield  {journal} {\bibinfo  {journal} {Astrophys. Space Sci. Proc.}\
  }\textbf {\bibinfo {volume} {38}},\ \bibinfo {pages} {107} (\bibinfo {year}
  {2014})},\ \Eprint {http://arxiv.org/abs/1302.0903} {arXiv:1302.0903
  [astro-ph.CO]} \BibitemShut {NoStop}%
\bibitem [{\citenamefont {Marsh}(2016)}]{Marsh:2015xka}%
  \BibitemOpen
  \bibfield  {author} {\bibinfo {author} {\bibfnamefont {D.~J.~E.}\
  \bibnamefont {Marsh}},\ }\href {\doibase 10.1016/j.physrep.2016.06.005}
  {\bibfield  {journal} {\bibinfo  {journal} {Phys. Rept.}\ }\textbf {\bibinfo
  {volume} {643}},\ \bibinfo {pages} {1} (\bibinfo {year} {2016})},\ \Eprint
  {http://arxiv.org/abs/1510.07633} {arXiv:1510.07633 [astro-ph.CO]}
  \BibitemShut {NoStop}%
\bibitem [{\citenamefont {Hui}\ \emph {et~al.}(2017)\citenamefont {Hui},
  \citenamefont {Ostriker}, \citenamefont {Tremaine},\ and\ \citenamefont
  {Witten}}]{Hui:2016ltb}%
  \BibitemOpen
  \bibfield  {author} {\bibinfo {author} {\bibfnamefont {L.}~\bibnamefont
  {Hui}}, \bibinfo {author} {\bibfnamefont {J.~P.}\ \bibnamefont {Ostriker}},
  \bibinfo {author} {\bibfnamefont {S.}~\bibnamefont {Tremaine}}, \ and\
  \bibinfo {author} {\bibfnamefont {E.}~\bibnamefont {Witten}},\ }\href
  {\doibase 10.1103/PhysRevD.95.043541} {\bibfield  {journal} {\bibinfo
  {journal} {Phys. Rev.}\ }\textbf {\bibinfo {volume} {D95}},\ \bibinfo {pages}
  {043541} (\bibinfo {year} {2017})},\ \Eprint
  {http://arxiv.org/abs/1610.08297} {arXiv:1610.08297 [astro-ph.CO]}
  \BibitemShut {NoStop}%
\bibitem [{\citenamefont {Ure\~na L\'opez}(2019)}]{Urena-Lopez:2019kud}%
  \BibitemOpen
  \bibfield  {author} {\bibinfo {author} {\bibfnamefont {L.~A.}\ \bibnamefont
  {Ure\~na L\'opez}},\ }\href {\doibase 10.3389/fspas.2019.00047} {\bibfield
  {journal} {\bibinfo  {journal} {Front. Astron. Space Sci.}\ }\textbf
  {\bibinfo {volume} {6}},\ \bibinfo {pages} {47} (\bibinfo {year}
  {2019})}\BibitemShut {NoStop}%
\bibitem [{\citenamefont {Seidel}\ and\ \citenamefont
  {Suen}(1990)}]{Seidel:1990jh}%
  \BibitemOpen
  \bibfield  {author} {\bibinfo {author} {\bibfnamefont {E.}~\bibnamefont
  {Seidel}}\ and\ \bibinfo {author} {\bibfnamefont {W.-M.}\ \bibnamefont
  {Suen}},\ }\href {\doibase 10.1103/PhysRevD.42.384} {\bibfield  {journal}
  {\bibinfo  {journal} {Phys. Rev.}\ }\textbf {\bibinfo {volume} {D42}},\
  \bibinfo {pages} {384} (\bibinfo {year} {1990})}\BibitemShut {NoStop}%
\bibitem [{\citenamefont {Guzman}\ and\ \citenamefont
  {Urena-Lopez}(2004)}]{Guzman:2004wj}%
  \BibitemOpen
  \bibfield  {author} {\bibinfo {author} {\bibfnamefont {F.~S.}\ \bibnamefont
  {Guzman}}\ and\ \bibinfo {author} {\bibfnamefont {L.~A.}\ \bibnamefont
  {Urena-Lopez}},\ }\href {\doibase 10.1103/PhysRevD.69.124033} {\bibfield
  {journal} {\bibinfo  {journal} {Phys. Rev.}\ }\textbf {\bibinfo {volume}
  {D69}},\ \bibinfo {pages} {124033} (\bibinfo {year} {2004})},\ \Eprint
  {http://arxiv.org/abs/gr-qc/0404014} {arXiv:gr-qc/0404014 [gr-qc]}
  \BibitemShut {NoStop}%
\bibitem [{\citenamefont {Hu}\ \emph {et~al.}(2000)\citenamefont {Hu},
  \citenamefont {Barkana},\ and\ \citenamefont {Gruzinov}}]{Hu:2000ke}%
  \BibitemOpen
  \bibfield  {author} {\bibinfo {author} {\bibfnamefont {W.}~\bibnamefont
  {Hu}}, \bibinfo {author} {\bibfnamefont {R.}~\bibnamefont {Barkana}}, \ and\
  \bibinfo {author} {\bibfnamefont {A.}~\bibnamefont {Gruzinov}},\ }\href
  {\doibase 10.1103/PhysRevLett.85.1158} {\bibfield  {journal} {\bibinfo
  {journal} {Phys. Rev. Lett.}\ }\textbf {\bibinfo {volume} {85}},\ \bibinfo
  {pages} {1158} (\bibinfo {year} {2000})},\ \Eprint
  {http://arxiv.org/abs/astro-ph/0003365} {arXiv:astro-ph/0003365 [astro-ph]}
  \BibitemShut {NoStop}%
\bibitem [{\citenamefont {Sahni}\ and\ \citenamefont
  {Wang}(2000)}]{Sahni:1999qe}%
  \BibitemOpen
  \bibfield  {author} {\bibinfo {author} {\bibfnamefont {V.}~\bibnamefont
  {Sahni}}\ and\ \bibinfo {author} {\bibfnamefont {L.-M.}\ \bibnamefont
  {Wang}},\ }\href {\doibase 10.1103/PhysRevD.62.103517} {\bibfield  {journal}
  {\bibinfo  {journal} {Phys. Rev.}\ }\textbf {\bibinfo {volume} {D62}},\
  \bibinfo {pages} {103517} (\bibinfo {year} {2000})},\ \Eprint
  {http://arxiv.org/abs/astro-ph/9910097} {arXiv:astro-ph/9910097 [astro-ph]}
  \BibitemShut {NoStop}%
\bibitem [{\citenamefont {Matos}\ and\ \citenamefont
  {Urena-Lopez}(2000)}]{Matos:2000ng}%
  \BibitemOpen
  \bibfield  {author} {\bibinfo {author} {\bibfnamefont {T.}~\bibnamefont
  {Matos}}\ and\ \bibinfo {author} {\bibfnamefont {L.~A.}\ \bibnamefont
  {Urena-Lopez}},\ }\href {\doibase 10.1088/0264-9381/17/13/101} {\bibfield
  {journal} {\bibinfo  {journal} {Class. Quant. Grav.}\ }\textbf {\bibinfo
  {volume} {17}},\ \bibinfo {pages} {L75} (\bibinfo {year} {2000})},\ \Eprint
  {http://arxiv.org/abs/astro-ph/0004332} {arXiv:astro-ph/0004332 [astro-ph]}
  \BibitemShut {NoStop}%
\bibitem [{\citenamefont {Schive}\ \emph {et~al.}(2014)\citenamefont {Schive},
  \citenamefont {Liao}, \citenamefont {Woo}, \citenamefont {Wong},
  \citenamefont {Chiueh}, \citenamefont {Broadhurst},\ and\ \citenamefont
  {Hwang}}]{Schive:2014hza}%
  \BibitemOpen
  \bibfield  {author} {\bibinfo {author} {\bibfnamefont {H.-Y.}\ \bibnamefont
  {Schive}}, \bibinfo {author} {\bibfnamefont {M.-H.}\ \bibnamefont {Liao}},
  \bibinfo {author} {\bibfnamefont {T.-P.}\ \bibnamefont {Woo}}, \bibinfo
  {author} {\bibfnamefont {S.-K.}\ \bibnamefont {Wong}}, \bibinfo {author}
  {\bibfnamefont {T.}~\bibnamefont {Chiueh}}, \bibinfo {author} {\bibfnamefont
  {T.}~\bibnamefont {Broadhurst}}, \ and\ \bibinfo {author} {\bibfnamefont
  {W.~Y.~P.}\ \bibnamefont {Hwang}},\ }\href {\doibase
  10.1103/PhysRevLett.113.261302} {\bibfield  {journal} {\bibinfo  {journal}
  {Phys. Rev. Lett.}\ }\textbf {\bibinfo {volume} {113}},\ \bibinfo {pages}
  {261302} (\bibinfo {year} {2014})},\ \Eprint {http://arxiv.org/abs/1407.7762}
  {arXiv:1407.7762 [astro-ph.GA]} \BibitemShut {NoStop}%
\bibitem [{\citenamefont {Mocz}\ \emph {et~al.}(2017)\citenamefont {Mocz},
  \citenamefont {Vogelsberger}, \citenamefont {Robles}, \citenamefont {Zavala},
  \citenamefont {Boylan-Kolchin}, \citenamefont {Fialkov},\ and\ \citenamefont
  {Hernquist}}]{Mocz:2017wlg}%
  \BibitemOpen
  \bibfield  {author} {\bibinfo {author} {\bibfnamefont {P.}~\bibnamefont
  {Mocz}}, \bibinfo {author} {\bibfnamefont {M.}~\bibnamefont {Vogelsberger}},
  \bibinfo {author} {\bibfnamefont {V.~H.}\ \bibnamefont {Robles}}, \bibinfo
  {author} {\bibfnamefont {J.}~\bibnamefont {Zavala}}, \bibinfo {author}
  {\bibfnamefont {M.}~\bibnamefont {Boylan-Kolchin}}, \bibinfo {author}
  {\bibfnamefont {A.}~\bibnamefont {Fialkov}}, \ and\ \bibinfo {author}
  {\bibfnamefont {L.}~\bibnamefont {Hernquist}},\ }\href {\doibase
  10.1093/mnras/stx1887} {\bibfield  {journal} {\bibinfo  {journal} {Mon. Not.
  Roy. Astron. Soc.}\ }\textbf {\bibinfo {volume} {471}},\ \bibinfo {pages}
  {4559} (\bibinfo {year} {2017})},\ \Eprint {http://arxiv.org/abs/1705.05845}
  {arXiv:1705.05845 [astro-ph.CO]} \BibitemShut {NoStop}%
\bibitem [{\citenamefont {Mocz}\ \emph {et~al.}(2019)\citenamefont {Mocz} \emph
  {et~al.}}]{Mocz:2019uyd}%
  \BibitemOpen
  \bibfield  {author} {\bibinfo {author} {\bibfnamefont {P.}~\bibnamefont
  {Mocz}} \emph {et~al.},\ }\href@noop {} {\  (\bibinfo {year} {2019})},\
  \Eprint {http://arxiv.org/abs/1911.05746} {arXiv:1911.05746 [astro-ph.CO]}
  \BibitemShut {NoStop}%
\bibitem [{\citenamefont {{Mocz}}\ \emph {et~al.}(2019)\citenamefont {{Mocz}},
  \citenamefont {{Fialkov}}, \citenamefont {{Vogelsberger}}, \citenamefont
  {{Becerra}}, \citenamefont {{Amin}}, \citenamefont {{Bose}}, \citenamefont
  {{Boylan-Kolchin}}, \citenamefont {{Chavanis}}, \citenamefont {{Hernquist}},
  \citenamefont {{Lancaster}}, \citenamefont {{Marinacci}}, \citenamefont
  {{Robles}},\ and\ \citenamefont {{Zavala}}}]{2019PhRvL.123n1301M}%
  \BibitemOpen
  \bibfield  {author} {\bibinfo {author} {\bibfnamefont {P.}~\bibnamefont
  {{Mocz}}}, \bibinfo {author} {\bibfnamefont {A.}~\bibnamefont {{Fialkov}}},
  \bibinfo {author} {\bibfnamefont {M.}~\bibnamefont {{Vogelsberger}}},
  \bibinfo {author} {\bibfnamefont {F.}~\bibnamefont {{Becerra}}}, \bibinfo
  {author} {\bibfnamefont {M.~A.}\ \bibnamefont {{Amin}}}, \bibinfo {author}
  {\bibfnamefont {S.}~\bibnamefont {{Bose}}}, \bibinfo {author} {\bibfnamefont
  {M.}~\bibnamefont {{Boylan-Kolchin}}}, \bibinfo {author} {\bibfnamefont
  {P.-H.}\ \bibnamefont {{Chavanis}}}, \bibinfo {author} {\bibfnamefont
  {L.}~\bibnamefont {{Hernquist}}}, \bibinfo {author} {\bibfnamefont
  {L.}~\bibnamefont {{Lancaster}}}, \bibinfo {author} {\bibfnamefont
  {F.}~\bibnamefont {{Marinacci}}}, \bibinfo {author} {\bibfnamefont {V.~H.}\
  \bibnamefont {{Robles}}}, \ and\ \bibinfo {author} {\bibfnamefont
  {J.}~\bibnamefont {{Zavala}}},\ }\href {\doibase
  10.1103/PhysRevLett.123.141301} {\bibfield  {journal} {\bibinfo  {journal}
  {\prl}\ }\textbf {\bibinfo {volume} {123}},\ \bibinfo {eid} {141301}
  (\bibinfo {year} {2019})},\ \Eprint {http://arxiv.org/abs/1910.01653}
  {arXiv:1910.01653 [astro-ph.GA]} \BibitemShut {NoStop}%
\bibitem [{\citenamefont {Matos}\ and\ \citenamefont
  {Urena-Lopez}(2007)}]{Matos:2007zza}%
  \BibitemOpen
  \bibfield  {author} {\bibinfo {author} {\bibfnamefont {T.}~\bibnamefont
  {Matos}}\ and\ \bibinfo {author} {\bibfnamefont {L.~A.}\ \bibnamefont
  {Urena-Lopez}},\ }\href {\doibase 10.1007/s10714-007-0470-y} {\bibfield
  {journal} {\bibinfo  {journal} {Gen. Rel. Grav.}\ }\textbf {\bibinfo {volume}
  {39}},\ \bibinfo {pages} {1279} (\bibinfo {year} {2007})}\BibitemShut
  {NoStop}%
\bibitem [{\citenamefont {Bernal}\ \emph {et~al.}(2010)\citenamefont {Bernal},
  \citenamefont {Barranco}, \citenamefont {Alic},\ and\ \citenamefont
  {Palenzuela}}]{Bernal:2009zy}%
  \BibitemOpen
  \bibfield  {author} {\bibinfo {author} {\bibfnamefont {A.}~\bibnamefont
  {Bernal}}, \bibinfo {author} {\bibfnamefont {J.}~\bibnamefont {Barranco}},
  \bibinfo {author} {\bibfnamefont {D.}~\bibnamefont {Alic}}, \ and\ \bibinfo
  {author} {\bibfnamefont {C.}~\bibnamefont {Palenzuela}},\ }\href {\doibase
  10.1103/PhysRevD.81.044031} {\bibfield  {journal} {\bibinfo  {journal} {Phys.
  Rev.}\ }\textbf {\bibinfo {volume} {D81}},\ \bibinfo {pages} {044031}
  (\bibinfo {year} {2010})},\ \Eprint {http://arxiv.org/abs/0908.2435}
  {arXiv:0908.2435 [gr-qc]} \BibitemShut {NoStop}%
\bibitem [{\citenamefont {Urena-Lopez}\ and\ \citenamefont
  {Bernal}(2010)}]{UrenaLopez:2010ur}%
  \BibitemOpen
  \bibfield  {author} {\bibinfo {author} {\bibfnamefont {L.~A.}\ \bibnamefont
  {Urena-Lopez}}\ and\ \bibinfo {author} {\bibfnamefont {A.}~\bibnamefont
  {Bernal}},\ }\href {\doibase 10.1103/PhysRevD.82.123535} {\bibfield
  {journal} {\bibinfo  {journal} {Phys. Rev.}\ }\textbf {\bibinfo {volume}
  {D82}},\ \bibinfo {pages} {123535} (\bibinfo {year} {2010})},\ \Eprint
  {http://arxiv.org/abs/1008.1231} {arXiv:1008.1231 [gr-qc]} \BibitemShut
  {NoStop}%
\bibitem [{\citenamefont {Li}\ \emph {et~al.}(2019)\citenamefont {Li},
  \citenamefont {Sun}, \citenamefont {Hu}, \citenamefont {Song},\ and\
  \citenamefont {Wang}}]{Li:2019mlk}%
  \BibitemOpen
  \bibfield  {author} {\bibinfo {author} {\bibfnamefont {H.-B.}\ \bibnamefont
  {Li}}, \bibinfo {author} {\bibfnamefont {S.}~\bibnamefont {Sun}}, \bibinfo
  {author} {\bibfnamefont {T.-T.}\ \bibnamefont {Hu}}, \bibinfo {author}
  {\bibfnamefont {Y.}~\bibnamefont {Song}}, \ and\ \bibinfo {author}
  {\bibfnamefont {Y.-Q.}\ \bibnamefont {Wang}},\ }\href@noop {} {\  (\bibinfo
  {year} {2019})},\ \Eprint {http://arxiv.org/abs/1906.00420} {arXiv:1906.00420
  [gr-qc]} \BibitemShut {NoStop}%
\bibitem [{\citenamefont {Silveira}\ and\ \citenamefont
  {de~Sousa}(1995)}]{Silveira:1995dh}%
  \BibitemOpen
  \bibfield  {author} {\bibinfo {author} {\bibfnamefont {V.}~\bibnamefont
  {Silveira}}\ and\ \bibinfo {author} {\bibfnamefont {C.~M.~G.}\ \bibnamefont
  {de~Sousa}},\ }\href {\doibase 10.1103/PhysRevD.52.5724} {\bibfield
  {journal} {\bibinfo  {journal} {Phys. Rev.}\ }\textbf {\bibinfo {volume}
  {D52}},\ \bibinfo {pages} {5724} (\bibinfo {year} {1995})},\ \Eprint
  {http://arxiv.org/abs/astro-ph/9508034} {arXiv:astro-ph/9508034 [astro-ph]}
  \BibitemShut {NoStop}%
\bibitem [{\citenamefont {Hertzberg}\ and\ \citenamefont
  {Schiappacasse}(2018)}]{Hertzberg:2018lmt}%
  \BibitemOpen
  \bibfield  {author} {\bibinfo {author} {\bibfnamefont {M.~P.}\ \bibnamefont
  {Hertzberg}}\ and\ \bibinfo {author} {\bibfnamefont {E.~D.}\ \bibnamefont
  {Schiappacasse}},\ }\href {\doibase 10.1088/1475-7516/2018/08/028} {\bibfield
   {journal} {\bibinfo  {journal} {JCAP}\ }\textbf {\bibinfo {volume} {1808}},\
  \bibinfo {pages} {028} (\bibinfo {year} {2018})},\ \Eprint
  {http://arxiv.org/abs/1804.07255} {arXiv:1804.07255 [hep-ph]} \BibitemShut
  {NoStop}%
\bibitem [{\citenamefont {Davidson}\ and\ \citenamefont
  {Schwetz}(2016)}]{Davidson:2016uok}%
  \BibitemOpen
  \bibfield  {author} {\bibinfo {author} {\bibfnamefont {S.}~\bibnamefont
  {Davidson}}\ and\ \bibinfo {author} {\bibfnamefont {T.}~\bibnamefont
  {Schwetz}},\ }\href {\doibase 10.1103/PhysRevD.93.123509} {\bibfield
  {journal} {\bibinfo  {journal} {Phys. Rev.}\ }\textbf {\bibinfo {volume}
  {D93}},\ \bibinfo {pages} {123509} (\bibinfo {year} {2016})},\ \Eprint
  {http://arxiv.org/abs/1603.04249} {arXiv:1603.04249 [astro-ph.CO]}
  \BibitemShut {NoStop}%
\bibitem [{\citenamefont {S{\'{e}}billeau}(1998)}]{Sebilleau1998}%
  \BibitemOpen
  \bibfield  {author} {\bibinfo {author} {\bibfnamefont {D.}~\bibnamefont
  {S{\'{e}}billeau}},\ }\href {\doibase 10.1088/0305-4470/31/34/017} {\bibfield
   {journal} {\bibinfo  {journal} {Journal of Physics A: Mathematical and
  General}\ }\textbf {\bibinfo {volume} {31}},\ \bibinfo {pages} {7157}
  (\bibinfo {year} {1998})}\BibitemShut {NoStop}%
\bibitem [{\citenamefont {Guzman}\ and\ \citenamefont
  {Urena-Lopez}(2006)}]{Guzman:2006yc}%
  \BibitemOpen
  \bibfield  {author} {\bibinfo {author} {\bibfnamefont {F.~S.}\ \bibnamefont
  {Guzman}}\ and\ \bibinfo {author} {\bibfnamefont {L.~A.}\ \bibnamefont
  {Urena-Lopez}},\ }\href {\doibase 10.1086/504508} {\bibfield  {journal}
  {\bibinfo  {journal} {Astrophys. J.}\ }\textbf {\bibinfo {volume} {645}},\
  \bibinfo {pages} {814} (\bibinfo {year} {2006})},\ \Eprint
  {http://arxiv.org/abs/astro-ph/0603613} {arXiv:astro-ph/0603613 [astro-ph]}
  \BibitemShut {NoStop}%
\bibitem [{\citenamefont {Olabarrieta}\ \emph {et~al.}(2007)\citenamefont
  {Olabarrieta}, \citenamefont {Ventrella}, \citenamefont {Choptuik},\ and\
  \citenamefont {Unruh}}]{Olabarrieta:2007di}%
  \BibitemOpen
  \bibfield  {author} {\bibinfo {author} {\bibfnamefont {I.}~\bibnamefont
  {Olabarrieta}}, \bibinfo {author} {\bibfnamefont {J.~F.}\ \bibnamefont
  {Ventrella}}, \bibinfo {author} {\bibfnamefont {M.~W.}\ \bibnamefont
  {Choptuik}}, \ and\ \bibinfo {author} {\bibfnamefont {W.~G.}\ \bibnamefont
  {Unruh}},\ }\href {\doibase 10.1103/PhysRevD.76.124014} {\bibfield  {journal}
  {\bibinfo  {journal} {Phys. Rev.}\ }\textbf {\bibinfo {volume} {D76}},\
  \bibinfo {pages} {124014} (\bibinfo {year} {2007})},\ \Eprint
  {http://arxiv.org/abs/0708.0513} {arXiv:0708.0513 [gr-qc]} \BibitemShut
  {NoStop}%
\bibitem [{\citenamefont {Alcubierre}\ \emph {et~al.}(2018)\citenamefont
  {Alcubierre}, \citenamefont {Barranco}, \citenamefont {Bernal}, \citenamefont
  {Degollado}, \citenamefont {Diez-Tejedor}, \citenamefont {Megevand},
  \citenamefont {Nunez},\ and\ \citenamefont {Sarbach}}]{Alcubierre:2018ahf}%
  \BibitemOpen
  \bibfield  {author} {\bibinfo {author} {\bibfnamefont {M.}~\bibnamefont
  {Alcubierre}}, \bibinfo {author} {\bibfnamefont {J.}~\bibnamefont
  {Barranco}}, \bibinfo {author} {\bibfnamefont {A.}~\bibnamefont {Bernal}},
  \bibinfo {author} {\bibfnamefont {J.~C.}\ \bibnamefont {Degollado}}, \bibinfo
  {author} {\bibfnamefont {A.}~\bibnamefont {Diez-Tejedor}}, \bibinfo {author}
  {\bibfnamefont {M.}~\bibnamefont {Megevand}}, \bibinfo {author}
  {\bibfnamefont {D.}~\bibnamefont {Nunez}}, \ and\ \bibinfo {author}
  {\bibfnamefont {O.}~\bibnamefont {Sarbach}},\ }\href {\doibase
  10.1088/1361-6382/aadcb6} {\bibfield  {journal} {\bibinfo  {journal} {Class.
  Quant. Grav.}\ }\textbf {\bibinfo {volume} {35}},\ \bibinfo {pages} {19LT01}
  (\bibinfo {year} {2018})},\ \Eprint {http://arxiv.org/abs/1805.11488}
  {arXiv:1805.11488 [gr-qc]} \BibitemShut {NoStop}%
\bibitem [{\citenamefont {Alcubierre}\ \emph {et~al.}(2019)\citenamefont
  {Alcubierre}, \citenamefont {Barranco}, \citenamefont {Bernal}, \citenamefont
  {Degollado}, \citenamefont {Diez-Tejedor}, \citenamefont {Megevand},
  \citenamefont {Núñez},\ and\ \citenamefont {Sarbach}}]{Alcubierre:2019qnh}%
  \BibitemOpen
  \bibfield  {author} {\bibinfo {author} {\bibfnamefont {M.}~\bibnamefont
  {Alcubierre}}, \bibinfo {author} {\bibfnamefont {J.}~\bibnamefont
  {Barranco}}, \bibinfo {author} {\bibfnamefont {A.}~\bibnamefont {Bernal}},
  \bibinfo {author} {\bibfnamefont {J.~C.}\ \bibnamefont {Degollado}}, \bibinfo
  {author} {\bibfnamefont {A.}~\bibnamefont {Diez-Tejedor}}, \bibinfo {author}
  {\bibfnamefont {M.}~\bibnamefont {Megevand}}, \bibinfo {author}
  {\bibfnamefont {D.}~\bibnamefont {Núñez}}, \ and\ \bibinfo {author}
  {\bibfnamefont {O.}~\bibnamefont {Sarbach}},\ }\href {\doibase
  10.1088/1361-6382/ab4726} {\bibfield  {journal} {\bibinfo  {journal} {Class.
  Quant. Grav.}\ }\textbf {\bibinfo {volume} {36}},\ \bibinfo {pages} {215013}
  (\bibinfo {year} {2019})},\ \Eprint {http://arxiv.org/abs/1906.08959}
  {arXiv:1906.08959 [gr-qc]} \BibitemShut {NoStop}%
\bibitem [{\citenamefont {{Rindler-Daller}}\ and\ \citenamefont
  {{Shapiro}}(2012)}]{2012MNRAS.422..135R}%
  \BibitemOpen
  \bibfield  {author} {\bibinfo {author} {\bibfnamefont {T.}~\bibnamefont
  {{Rindler-Daller}}}\ and\ \bibinfo {author} {\bibfnamefont {P.~R.}\
  \bibnamefont {{Shapiro}}},\ }\href {\doibase
  10.1111/j.1365-2966.2012.20588.x} {\bibfield  {journal} {\bibinfo  {journal}
  {MNRAS}\ }\textbf {\bibinfo {volume} {422}},\ \bibinfo {pages} {135}
  (\bibinfo {year} {2012})},\ \Eprint {http://arxiv.org/abs/1106.1256}
  {arXiv:1106.1256 [astro-ph.CO]} \BibitemShut {NoStop}%
\bibitem [{\citenamefont {Guzmán}\ \emph {et~al.}(2014)\citenamefont
  {Guzmán}, \citenamefont {Lora-Clavijo}, \citenamefont {González-Avilés},\
  and\ \citenamefont {Rivera-Paleo}}]{Guzman:2013rua}%
  \BibitemOpen
  \bibfield  {author} {\bibinfo {author} {\bibfnamefont {F.~S.}\ \bibnamefont
  {Guzmán}}, \bibinfo {author} {\bibfnamefont {F.~D.}\ \bibnamefont
  {Lora-Clavijo}}, \bibinfo {author} {\bibfnamefont {J.~J.}\ \bibnamefont
  {González-Avilés}}, \ and\ \bibinfo {author} {\bibfnamefont {F.~J.}\
  \bibnamefont {Rivera-Paleo}},\ }\href {\doibase 10.1103/PhysRevD.89.063507}
  {\bibfield  {journal} {\bibinfo  {journal} {Phys. Rev.}\ }\textbf {\bibinfo
  {volume} {D89}},\ \bibinfo {pages} {063507} (\bibinfo {year} {2014})},\
  \Eprint {http://arxiv.org/abs/1310.3909} {arXiv:1310.3909 [astro-ph.CO]}
  \BibitemShut {NoStop}%
\bibitem [{\citenamefont {http://www.ifm.umich.mx/
  \~{}guzman/SupplementalMultistates/}()}]{supplemental}%
  \BibitemOpen
  \bibfield  {author} {\bibinfo {author} {\bibnamefont
  {http://www.ifm.umich.mx/ \~{}guzman/SupplementalMultistates/}},\ }\href@noop
  {} {\ }\BibitemShut {NoStop}%
\bibitem [{\citenamefont {Guzm\'an}\ and\ \citenamefont
  {Avilez}(2018)}]{PhysRevD.97.116003}%
  \BibitemOpen
  \bibfield  {author} {\bibinfo {author} {\bibfnamefont {F.~S.}\ \bibnamefont
  {Guzm\'an}}\ and\ \bibinfo {author} {\bibfnamefont {A.~A.}\ \bibnamefont
  {Avilez}},\ }\href {\doibase 10.1103/PhysRevD.97.116003} {\bibfield
  {journal} {\bibinfo  {journal} {Phys. Rev. D}\ }\textbf {\bibinfo {volume}
  {97}},\ \bibinfo {pages} {116003} (\bibinfo {year} {2018})}\BibitemShut
  {NoStop}%
\bibitem [{\citenamefont {Matos}\ \emph {et~al.}()\citenamefont {Matos},
  \citenamefont {Sol\'is-L\'opez}, \citenamefont {Guzm\'an}, \citenamefont
  {Robles},\ and\ \citenamefont {Ure\~na L\'opez}}]{Satellites}%
  \BibitemOpen
  \bibfield  {author} {\bibinfo {author} {\bibfnamefont {T.}~\bibnamefont
  {Matos}}, \bibinfo {author} {\bibfnamefont {J.}~\bibnamefont
  {Sol\'is-L\'opez}}, \bibinfo {author} {\bibfnamefont {F.~S.}\ \bibnamefont
  {Guzm\'an}}, \bibinfo {author} {\bibfnamefont {V.~H.}\ \bibnamefont
  {Robles}}, \ and\ \bibinfo {author} {\bibfnamefont {L.~A.}\ \bibnamefont
  {Ure\~na L\'opez}},\ }\href@noop {} {\ }\Eprint
  {http://arxiv.org/abs/1912.09660} {arXiv:1912.09660 [astro-ph]} \BibitemShut
  {NoStop}%
\end{thebibliography}%

%

\end{document}